\documentclass[11pt,paper]{article}
\usepackage{jheppub}
\usepackage[usenames,dvipsnames,table]{xcolor}
\usepackage{graphicx,amsmath,amssymb,multirow,array,bm,mathrsfs}
\usepackage{epsf,amsfonts}
\usepackage[numbers,sort&compress]{natbib}

\usepackage{slashed}

\newcommand{\fixform}[1]{\texorpdfstring{#1}{~}}

\newcommand{\beq}{\begin{equation}}
\newcommand{\eeq}{\end{equation}}
\newcommand{\bea}{\begin{eqnarray}}
\newcommand{\eea}{\end{eqnarray}}

\newcommand{\prn}[1]{\left ( #1 \right )}
\newcommand{\brk}[1]{\left [ #1 \right ]}
\newcommand{\bigbr}[1]{\Bigl\{ #1 \Bigr\} }

\newcommand{\half}{\frac{1}{2}}
\newcommand{\quarter}{\frac{1}{4}}

\newcommand{\tr}{\mbox{tr}}

\newcommand{\form}[1]{\bm{#1}}
\newcommand{\diffF}{{\delta_\chi}}
\newcommand{\delnc}{{\delta_\chi^{\text{non-cov}}}}
\newcommand{\ic}{\form{i}}
\newcommand{\hodge}{{}^\star}
\newcommand{\lieD}{\pounds}
\newcommand{\hodgeCFT}{ {\hodge}^{\text{\tiny{CFT}}}}

\newcommand{\fA}{\form{A}}

\newcommand{\fF}{\form{F}}

\newcommand{\fGamma}{\form{\Gamma}}

\newcommand{\fR}{\form{R}}

\newcommand{\Sp}{\Sigma}

\newcommand{\fP}{{\form{\mathcal{P}}}}

\newcommand{\ICS}{{\form{I}}_{CS}}

\newcommand{\JH}{\mathrm{J}_{H}}

\newcommand{\fJH}{\form{\mathrm{J}}_{H}}

\newcommand{\SpH}{\mathrm{\Sp}_{H}}

\newcommand{\fSpH}{\form{\mathrm{\Sp}}_{H}}

\newcommand{\THall}{\mathrm{T}_{H}}

\newcommand{\sHFF}{\bar{\sigma}^{FF}_{_H}}
\newcommand{\sHFR}{\bar{\sigma}^{FR}_{_H}}
\newcommand{\sHRF}{\bar{\sigma}^{RF}_{_H}}
\newcommand{\sHRR}{\bar{\sigma}^{RR}_{_H}}

\newcommand{\fsHFF}{\form{\sigma}^{FF}_{_H}}
\newcommand{\fsHFR}{\form{\sigma}^{FR}_{_H}}
\newcommand{\fsHRF}{\form{\sigma}^{RF}_{_H}}
\newcommand{\fsHRR}{\form{\sigma}^{RR}_{_H}}

\newcommand{\eom}[1]{\slashed{\delta}_{#1}\overline{\mathcal{E}}}
\newcommand{\feom}[1]{\slashed{\delta}_{#1}\form{\mathcal{E}}}
\newcommand{\eomChi}{\slashed{\delta}_\chi\overline{\mathcal{E}}}
\newcommand{\PSympl}{\slashed{\delta}^2\overline{\Omega}_{_{\text{PSympl}}}}
\newcommand{\fPSympl}{\slashed{\delta}^2\form{\Omega}_{\text{PSympl}}}
\newcommand{\PSymChi}{\slashed{\delta}\slashed{\delta}_\chi\overline{\Omega}_{_{\text{PSympl}}}}
\newcommand{\fPSymChi}{\slashed{\delta}\slashed{\delta}_\chi\form{\Omega}_{\text{PSympl}}}
\newcommand{\PSymplPot}[1]{\slashed{\delta}_{#1}\overline{\varTheta}_{_{\text{PSympl}}}}
\newcommand{\fPSymplPot}[1]{\slashed{\delta}_{#1}\form{\varTheta}_{\text{PSympl}}}
\newcommand{\PSymplPotChi}{\slashed{\delta}_\chi\overline{\varTheta}_{_{\text{PSympl}}}}
\newcommand{\Lag}{\overline{L}}
\newcommand{\fLag}{\form{L}}
\newcommand{\QNoether}{\slashed{\delta}\overline{Q}_{_{\text{Noether}}}}
\newcommand{\fQNoether}{\slashed{\delta}\form{Q}_{_{\text{Noether}}}}
\newcommand{\Komar}{\overline{\mathcal{K}}_\chi}
\newcommand{\fKomar}{\form{\mathcal{K}}_\chi}

\newcommand{\fXi}{\form{\Xi}_\chi}
\newcommand{\fSigma}{\slashed{\delta}\form{\Sigma}_\chi}

\newcommand{\fV}{\form{V}}
\newcommand{\fY}{\form{Y}}
\newcommand{\fZ}{\slashed{\delta}\form{Z}}

\newcommand{\N}{\mathrm{N}}
\newcommand{\fN}{\form{\N}}

\newcommand{\fPCFT}{\fP_{CFT}}

\newcommand{\GN}{G_{_N}}
\newcommand{\cc}{\Lambda_{_{cc}}}
\newcommand{\gYM}{g_{_{EM}}}
\newcommand{\TMax}{\mathrm{T}_{_M}}

\newcommand{\CFT}{{\text{\tiny{CFT}}}}

\newcommand{\xiH}{\bm{\beta}}
\newcommand{\LambdaH}{\bm{\Lambda_{_\beta}}}

\title{Covariant Noether Charge for Higher Dimensional Chern-Simons Terms}

\author[a]{Tatsuo Azeyanagi,}
\author[b]{R. Loganayagam,}
\author[c]{Gim Seng Ng}
\author[c, d]{and Maria J. Rodriguez}

\affiliation[a]{
D\'{e}partement de Physique, Ecole Normale Sup\'{e}rieure, CNRS, 24 rue Lhomond,
75005 Paris, France.}
\affiliation[b]{School of Natural Sciences, Institute for Advanced Study, Princeton, NJ 08540, USA.}
\affiliation[c]{Center for the Fundamental Laws of Nature, Harvard University, Cambridge, MA 02138, USA.}
\affiliation[d]{Institut de Physique Th\'eorique, CEA Saclay, CNRS URA 2306 , F-91191 Gif-sur-Yvette, France.}

\emailAdd{tatsuo.azeyanagi@phys.ens.fr}
\emailAdd{nayagam@ias.edu}
\emailAdd{nggimseng@post.harvard.edu}
\emailAdd{mjrodri@physics.harvard.edu}

\abstract{
We construct a manifestly covariant differential Noether charge for theories with Chern-Simons terms in higher dimensional spacetimes.
 This is in contrast to Tachikawa's extension of the standard Lee-Iyer-Wald formalism which results in a non-covariant differential Noether charge for Chern-Simons terms.
On a bifurcation surface, our differential Noether charge integrates to the Wald-like entropy formula proposed by Tachikawa in [\href{http://arxiv.org/abs/hep-th/0611141}{\tt arXiv:hep-th/0611141}].
}
\begin{document}
\maketitle

\section{Introduction}
One of the remarkable aspects of gravity is the fact that classical black hole solutions have a finite entropy. The question 
of how this entropy is encoded in the geometry of the black hole solution is a longstanding problem which has 
motivated much recent research. In Einstein gravity, the answer to this question is given in terms of the 
celebrated Bekenstein-Hawking formula $S= \frac{A}{4\GN}$ which gives a simple way of reading off 
the entropy of a black hole solution from its horizon area. Despite a variety of efforts, a formula of such generality 
is not yet known for higher derivative gravity.

A formula applicable to specific limits is however known - an important progress in this direction is the Wald formula\cite{Jacobson:1993xs,Wald:1993nt,Jacobson:1993vj,Iyer:1994ys,Jacobson:1994qe,Jacobson:1995uq,Gao:2001ut} applicable to time-independent geometries which is constructed by demanding the first law of thermodynamics.\footnote{Whether Wald entropy obeys the second law of thermodynamics is however still an open question.} Wald gave a particular prescription in the context of the Noether procedure\footnote{By this we mean the collection of various formalisms which 
rely on some version of Noether charge - apart from the treatment by Lee-Iyer-Wald\cite{Lee:1990nz,Iyer:1994ys}, there are related
methods commonly attributed to Abbott-Deser-Tekin \cite{Abbott:1981ff,Deser:2002rt,Deser:2002jk,Deser:2003vh} and 
Barnich-Brandt-Comp\`{e}re\cite{Barnich:2001jy,Barnich:2007bf,Compere:2007az}.}  whereby he identified 
an appropriate Noether charge at the horizon as the entropy. The Wald formula has had many successes : microscopic computations of entropy (via say Sen's entropy function formalism
for extremal black holes \cite{Sen:2005wa,Sen:2007qy,Sahoo:2006vz}) reduce to Wald entropy in appropriate limits. Entanglement entropy computations in AdS/CFT exhibit Wald-like formula with corrections\cite{Camps:2013zua,Dong:2013qoa} thus 
giving a geometric realization of the interplay between thermal entropy and entanglement entropy. 
Attempts to generalize the Wald formula to time-dependent situations however runs into various ambiguities and the physical principle to resolve 
these ambiguities are still unknown. 

The main obstacle to using AdS/CFT to resolve these questions is the fact  that time-dependent entropies are difficult to compute even in field theory.  It thus seems essential  that we find simple time-dependent situations where we can study how entropy is geometrized in gravity. A simple situation which might be tractable is the entropy associated with anomalies in field theory. The robustness of anomalies could allow us to understand quantitatively the associated anomaly even in time-dependent cases\cite{Jensen:Nov2013}. AdS/CFT then maps this situation to the case of gravitational solutions in the presence of Chern-Simons terms. One thus hopes that understanding Wald-type entropy that arises from Chern-Simons terms might lead us to a better understanding of the geometric entropy and the way to generalize it.

The original derivation by Wald assumes covariant Lagrangians and hence excludes Chern-Simons terms. The
Lee-Iyer-Wald formalism for constructing Noether charge was later extended to theories with Chern-Simons terms by 
Tachikawa \cite{Tachikawa:2006sz} (this proposal was then worked out in detail by  
Bonora-Cvitan-Prester-Pallua-Smolic\cite{Bonora:2011gz}) which we will review when we compare with our results. This Tachikawa's extension, however, is not manifestly covariant and it runs 
aground with issues of covariance\cite{Bonora:2011gz}  in dimensions greater than three.\footnote{Most of the applications of Tachikawa's prescription has been for the pure gravitational Chern-Simons term in AdS$_3$ where the covariance of the final results can be easily 
demonstrated. In fact, the three dimensional gravitational Chern-Simons term has been widely studied\cite{Kraus:2005zm,Solodukhin:2005ah,Bouchareb:2007yx,Compere:2008cv,Miskovic:2009kr,Nazaroglu:2011zi,Kim:2013cor} in the context of topologically massive gravity.
See \cite{Mora:2006ka,Bonora:2011gz,Bonora:2011mf,Bonora:2012xv,Bonora:2012eb,Bonora:2013jca} also for discussions on higher dimensional Chern-Simons terms.
 }

In this work, we will trace these  issues to the use of a non-covariant pre-symplectic structure on the space of solutions. Our main motivation in this work is to demonstrate that, with higher dimensional  Chern-Simons terms, one can instead choose a  manifestly covariant  pre-symplectic structure and implement the  Noether procedure  in a manifestly covariant way.\footnote{We remind the reader that the issue of covariance of charges in the presence of Chern-Simons terms is often a subtle issue\cite{Marolf:2000cb}. What we are interested in roughly corresponds to what Marolf calls the `Maxwell charge'. From the dual CFT point of view, we want a Noether procedure that would  compute for us the covariant currents.} Using our pre-symplectic current we will  then re-derive the final entropy formula proposed in \cite{Tachikawa:2006sz} without having 
to choose special gauges/coordinates systems (as is necessary in the method described in \cite{Tachikawa:2006sz,Bonora:2011gz}).

In fact, this is a general lesson which underscores why Chern-Simons terms serve as stringent tests for any generalized entropy proposal : 
most constructions  and ideas about how the Wald formula should be generalized often do not work for Chern-Simons terms
because of  covariance issues. Our hope is that our analysis in this paper would help us tease out the essential 
features of the Noether procedure that survive this `Chern-Simons' test  so that we can be guided as to how we 
should go about generalizing it in time-dependent situations.

We will divide the rest of this introduction into four different subsections. In the subsection that follows we begin by introducing 
Einstein-Maxwell-Chern-Simons system \`{a} la  \cite{Azeyanagi:2013xea}. The aim is to introduce notation as 
well to present the reader with a specific context where our results can be used. In the subsequent subsection we
quickly introduce the essential  ideas of the Noether formalism that the reader would need to understand 
the third subsection summarizing of our results. In the final subsection, we provide the outline of this paper.

\subsection*{System under study}
A main motivation for this paper is  our recent work in \cite{Azeyanagi:2013xea} where using fluid/gravity correspondence, we 
constructed a class of AdS black hole  solutions for Einstein-Maxwell-Chern-Simons equations in AdS$_{2n+1}$.
That construction was in turn motivated by recent advances in the field theory side on how Lorentz anomalies
enter into hydrodynamics\cite{Landsteiner:2011cp, Landsteiner:2011iq, Loganayagam:2011mu,Loganayagam:2012pz, 
Banerjee:2012cr,Jensen:2012kj,Loganayagam:2012zg,Jensen:2013kka,Jensen:2013rga,Yee:2014dxa}.
Since we will develop our covariant prescription in the context of this system, we begin by reviewing it.

We consider the simplest class of gravitational systems in AdS$_{d+1}$ with Chern-Simons terms with an action
\begin{equation}\label{eq:action}
\begin{split}
\int d^{d+1} x \,\sqrt{-G} \brk{ \frac{1}{16\pi \GN} \prn{R-2\cc}
-\frac{1}{4\gYM^2}  F_{ab} F^{ab} }+  \int \ICS[\fA,\fF,\fGamma,\fR]\, , 
\end{split}
\end{equation}
where  the Chern-Simons part of the Lagrangian is denoted as $\ICS$ which is a $d+1$ form.
Since Chern-Simons terms are odd forms, this necessarily implies that $d=2n$ with 
$n$ an integer. This action then leads to the equations of motion :
\begin{equation}\label{eq:EinsteinMaxwellEq}
\begin{split}
R_{ab}-\frac{1}{2}\prn{R-2\cc} G_{ab} &=8\pi \GN  \brk{ (\TMax)_{ab} +  (\THall)_{ab} }\ ,\\
D^b F_{ab} &= \gYM^2 (\JH)_a\,,
\end{split}
\end{equation}
where $G_{ab}$ is an asymptotically AdS$_{d+1}$ metric with $d=2n$, $F_{ab}$ is 
the Maxwell field strength defined from the vector potential $A_a$ via $F_{ab} \equiv \partial_a A_b - \partial_b A_a$. All our expressions in this work equally well apply to the Yang-Mills system where $F_{ab} \equiv  \partial_a A_b - \partial_b A_a + [A_a,A_b]$ . With this in mind, we use $D_b$ to denote the gauge covariant derivative.
Here, $\GN$  and $\gYM$ are the Newton and Maxwell couplings 
respectively. The cosmological constant is taken to be negative and 
is given by $\cc\equiv  -d(d-1)/2$ where the AdS radius is set to one.

The Maxwell energy-momentum tensor $ (\TMax)^{ab}$ in the above equation takes the form 
\begin{equation}
\begin{split}
 (\TMax)^{ab}
&\equiv
 \frac{1}{\gYM^2}\ \brk{ F^{ac}F^b{}_c-\quarter G^{ab} F_{cd} F^{cd} }\, ,\\
\end{split}
\end{equation}
whereas $(\THall)_{ab}$ and $(\JH)_a$ are the energy-momentum tensor and the Maxwell charge current
obtained by varying the Chern-Simons part of the action. We will call these currents as Hall currents.
The bulk Hall currents  are more conveniently written in terms of the formal $(d+2)$-form $\fPCFT=d\ICS$, the anomaly polynomial of the dual CFT.  We note that the anomaly polynomial depends only on the
Maxwell field strength two-form  $\fF$ and the curvature two-form  $\fR^a{}_b$, both of which are covariant. On the other hand, the Chern-Simons form $\form{I}_{CS}$ depends on $\fF$ and $\fR^a{}_b$ as well as non-covariant quantities, i.e. the gauge field one-form $\form{A}$ or the connection one-form $\fGamma^a{}_b$\,.  
We define the spin Hall current $(\SpH)^{cb}{}_a$ and 
the charge Hall current $(\JH)^c$ corresponding to $\ICS$ as
\begin{equation}
\begin{split}\label{eq:defsHall}
(\hodge \fSpH)^b{}_a &\equiv (\SpH)^{cb}{}_a\  \hodge dx_c  
\equiv -2 \prn{\frac{\partial \fPCFT}{ \partial \fR^a{}_b}}\,, \\
\hodge \fJH &\equiv (\JH)^c\  \hodge dx_c  \equiv -\prn{\frac{\partial \fPCFT}{ \partial \fF}}\, .   \\
\end{split}
\end{equation}
By varying the Chern-Simons Lagrangian $\ICS$ with respect to the metric $G_{ab}$, we can obtain the energy-momentum tensor associated with the Hall current
(sometimes called the generalized Cotton tensor\cite{Solodukhin:2005ns})
which is written as $(\THall)^{ab}=  \nabla_c (\SpH)^{(ab)c}$.

In \cite{Azeyanagi:2013xea}, we found  charged rotating black hole solutions of this system of equations in a 
fluid/gravity expansion. In this paper, we will construct a Noether charge prescription which will allow us to
assign energy, charge and entropy for these solutions. While this paper deals with the formal aspects of this
construction including the crucial issue of covariance, in an accompanying paper \cite{Azeyanagi:2014} we utilize this construction 
to compute the charges and entropy of our solutions and match them against CFT predictions.

\subsection*{Noether formalism}

Let us begin by  reviewing how the Noether formalism allows us to compute energy, 
entropy etc. Since we will be discussing this formalism extensively in the main text (with an eye towards 
Chern-Simons terms), we will be necessarily brief just outlining the main ideas needed for the rest 
of this introduction. Further, we will phrase the formalism in a language well-adopted to AdS/CFT
and fluid/gravity correspondence. 

Associated with every diffeomorphism or gauge transformation parametrized by $\{\xi^a, \Lambda\}$,  there is a co-dimension two form $\fQNoether$ which is  linear in variations of AdS fields  : 
we will call it \textbf{the differential Noether charge}. 
The symbol $\slashed{\delta}$ denotes that it is linear in variations of the fields and that it is not necessarily 
an integrable variation, viz.,  in general,  $\fQNoether\neq \delta \form{Q}$ for any $\form{Q}$.

Further, the Noether formalism implies that the exterior derivative of the differential Noether charge
$d\fQNoether$  associated with a $\{\xi^a, \Lambda\}$ is proportional on-shell to Lie derivatives  of the fields
along that $\{\xi^a, \Lambda\}$. The tensor of proportionality is given by a co-dimension
one form called  \textbf{the pre-symplectic current} $\fPSympl$. The pre-symplectic current is proportional to the 
product of two field variations as its notation indicates and it is antisymmetric under the exchange of the field variations. 
We can then write $d\fQNoether= - \fPSymChi$ where the subscript $\chi$ indicates that the second variation has 
been converted into a Lie-derivative along $\{\xi^a, \Lambda\}$.

The differential Noether charge $\fQNoether$ when restricted to a hypersurface in AdS  
becomes a  co-dimension one form.  We first consider $\fQNoether$ associated with a diffeomorphism/gauge transformation 
$\{\xi^a, \Lambda\}$  which acts on the dual CFT as a symmetry transformation  $\{\xi_\CFT^\mu, \Lambda^\CFT\}$, i.e., 
$\{\xi^a, \Lambda\}$ fall off slowly enough near the boundary of AdS that they act non-trivially on the boundary. We have
\[ \{\xi^a, \Lambda\}|_{\infty} = \{\xi_\CFT^\mu, \Lambda^\CFT\}\ .\]
Here $|_\infty$ denotes that the evaluation is carried out at the boundary. 
The $\fQNoether$ of such a $\{\xi^a, \Lambda\}$ is then restricted to a radial slice 
near the boundary of AdS and evaluated on-shell, i.e., we evaluate it on a solution to the gravity equations with the field variations satisfying linearized equations. This on-shell differential Noether charge 
then encodes the information about the   energy-momentum  and charge differences in the 
neighborhood of the state under consideration.  More precisely, we have 
\begin{equation} \label{eq:delQTJ}
 \fQNoether |_{\infty}
 =  -\left[ \eta_{\nu\sigma} \xi_\CFT^\sigma \delta T_\CFT^{\mu\nu}
 +(\Lambda^\CFT+\xi_\CFT^\alpha A_\alpha^\CFT )\  \delta J^\mu_\CFT \right]
  \,\, \hodgeCFT dx_\mu\ + d(\ldots) \,,
 \end{equation} 
where $\{ T_\CFT^{\mu\nu}, J^\mu_\CFT \}$ are the (expectation values of) energy-momentum tensor and the charge current of the dual CFT,  $\{\eta_{\nu\sigma}, A_\alpha^\CFT \}$ are the corresponding metric/gauge field sources in the CFT and $\hodgeCFT$ represents the CFT Hodge-dual operator acting on forms.\footnote{For details regarding our conventions for differential forms, the reader can consult Appendix~\ref{ss:forms}.}  Here $\delta T_\CFT^{\mu\nu}$ for example, is to be understood as the difference in  (the expectation value of) energy-momentum tensor in the neighborhood of  the dual CFT state. The term $d(\ldots)$ at the  end of Eq.~\eqref{eq:delQTJ} indicates that Eq.~\eqref{eq:delQTJ} is supposed to be valid up to an addition of an exact form.

The essential insight due to Wald is that,  at least as far as time-independent solutions go, the same differential Noether 
charge for an appropriate $\{\xi^a, \Lambda\}$  evaluated at the horizon gives the entropy of the solution. To give a 
more precise statement, we begin with the time-like Killing symmetry/gauge transformation $\{\xiH^a, \LambdaH\}$
which leaves invariant the time-independent state under question. We will assume further that the black hole horizon 
is a Killing horizon for $\{\xiH^a, \LambdaH\}$  with $\xiH^a$ having a surface gravity normalized to $2\pi$. 
This implies that  $\xiH^a = 0$ at the bifurcation surface and
\begin{equation}
 \{\ G_{ab} \xiH^a \xiH^b   = 0 ,\quad \xiH^b \nabla_b \xiH^a = 2\pi \xiH^a,\quad\LambdaH + \xiH^a A_a = 0 \ \} \qquad\text{at the horizon},
\end{equation}
where $\{G_{ab},A_a\} $ represent the bulk metric/gauge field. 
Roughly, one can think of $\xiH^a$ as the 
`inverse temperature' vector - more precisely its norm gives the  length of the thermal circle in the corresponding
Euclidean solution. Thus, it is null at the horizon where the Euclidean solution caps off and near the AdS boundary it
is a time-like vector whose norm gives the inverse temperature of the dual CFT.  

For a time-independent solution in fluid/gravity correspondence, $\{\xiH^a, \LambdaH\}$ can be computed in a boundary derivative expansion. 
In the usual ingoing Eddington-Finkelstein coordinates used in fluid/gravity correspondence, we get the  expansion :
\[ \{\xiH^a, \LambdaH + \xiH^b A_b \}|_{\infty} = \{\xiH_\CFT^\mu, \LambdaH^\CFT+\xiH_\CFT^\alpha A_\alpha^\CFT\} = \left\{\frac{u^\mu}{T}, \frac{\mu}{T}\right\} +\ldots , \]
where $\{u^\mu, T,\mu \}$ are the velocity, temperature and chemical potential fields of the CFT fluid. Wald argued 
that the on-shell  $\fQNoether$ corresponding to such a $\{\xiH^a, \LambdaH\}$ gives the entropy of the solution when restricted to the horizon,viz.,
\begin{equation}\label{eq:delQJS}
 \fQNoether |_{hor}= \delta J_{S,\CFT}^\mu\,   \hodgeCFT dx_\mu\, + d(\ldots)\,,
\end{equation} 
where $ J_{S,\CFT}^\mu$ is the entropy current of the dual CFT.  Here the symbol $|_{hor}$ represents an evaluation of $\fQNoether$ corresponding to  $\{\xiH^a, \LambdaH\}$  at the horizon on-shell, followed by a pullback of the answer to the boundary along ingoing null geodesics (in accordance with the usual fluid/gravity prescription for the CFT entropy current \cite{Bhattacharyya:2008xc}). This expression then provides us with a way of computing the entropy current for higher derivative fluid/gravity correspondence.

The advantage of assigning entropy via differential Noether charge is that the first law of thermodynamics follows immediately as a 
consequence of  the Noether formalism.  Since for time-independent solutions the differential Noether charge is closed on-shell, viz., 
 $d\fQNoether = 0$,   Eq.~\eqref{eq:delQJS} can equally well be evaluated near 
 the boundary of AdS. Using  Eq.~\eqref{eq:delQTJ} we can then write 
 \begin{equation}\label{eq:delQhorInf}
 \begin{split}
 \fQNoether |_{hor}&= \fQNoether |_{\infty} \\
 &= -\brk{ \eta_{\nu\sigma} \xiH_\CFT^\sigma \delta T_\CFT^{\mu\nu}
 +(\LambdaH^\CFT +\xiH_\CFT^\alpha A_\alpha^\CFT )\  \delta J^\mu_\CFT }
  \quad \hodgeCFT dx_\mu + d(\ldots) \,.
  \end{split}
\end{equation} 
Comparing equations Eq.~\eqref{eq:delQhorInf} against Eq.~\eqref{eq:delQJS}, we immediately get the CFT first law of thermodynamics  :
 \begin{equation}\label{eq:first law}
 \delta J_{S,\CFT}^\mu + \eta_{\nu\sigma} \xiH_\CFT^\sigma \delta T_\CFT^{\mu\nu}
 +(\LambdaH^\CFT +\xiH_\CFT^\alpha A_\alpha^\CFT )\  \delta J^\mu_\CFT  = 0 \, . 
 \end{equation} 

When the gravity Lagrangian $\Lag$ is manifestly covariant, i.e. if it does not contain Chern-Simons terms,
and if Eq.~\eqref{eq:delQJS} is integrated over the bifurcation surface,  we can remove the variations to write 
(by denoting $\Lag=\Lag_{cov}$ for later purpose)\cite{Wald:1993nt,Iyer:1994ys}
\begin{equation}\label{eq:Wald}
 S_{\text{Wald}} = \int_{Bif} 2\pi \varepsilon_b{}^a  \frac{\delta\Lag_{cov}}{\delta R^a{}_{bcd}} \varepsilon^{cd} = \int_{S_{\infty}} J_{S,\CFT}^\mu~  \hodgeCFT dx_\mu\, , 
\end{equation} 
where the left integral is over the bifurcation surface whereas the right integral is over a time slice 
in the CFT.\footnote{For a differential form $\fV$, 
the definition of $\overline{\form{V}}$ is given in Eq.~\eqref{eq:defofbarnotation}. }
Here $\varepsilon^{ab}$ is the binormal at the bifurcation surface defined via $\nabla_a \xiH^b |_{Bif}=  2\pi \varepsilon_a{}^b$ and  $\frac{\delta \Lag_{cov}}{\delta R^a_{bcd}}$ refers to a functional differentiation of the Lagrangian treating Riemann tensor as an independent
field . In time-independent solutions, the integral over the bifurcation surface can be replaced by a suitable integral over an arbitrary
time slice  of the horizon\cite{Jacobson:1993vj}.

Although this is the most common form of Wald entropy used in the literature, it is inapplicable precisely in the systems we are interested in, where  $\Lag$  contains Chern-Simons terms. For these systems, Tachikawa\cite{Tachikawa:2006sz} has proposed that
Eq.~\eqref{eq:Wald} be modified to
\begin{equation}\label{eq:WaldTachikawa}
\begin{split}
 S_{\text{Wald-Tachikawa}} &= \int_{Bif} 2\pi \varepsilon_b{}^a  \frac{\delta\Lag_{cov}}{\delta R^a{}_{bcd}} \varepsilon^{cd}  
 +  \int_{Bif}  \sum_{k=1}^\infty 8\pi k\ \fGamma_N (d\fGamma_N)^{2k-2} \frac{\partial \fPCFT}{\partial\ \text{tr} \fR^{2k} }\\
 &= \int_{S_{\infty}} J_{S,\CFT}^\mu\,\,  \hodgeCFT dx_\mu\, , 
 \end{split}
\end{equation} 
where $\Lag_{cov}$ is the covariant part of the gravity Lagrangian and $\fPCFT= d\ICS$ encodes the information about the Chern-Simons part.
Further, we have written the answer in terms of the normal bundle connection $ \fGamma_N $ and its curvature $ \fR_N = d\fGamma_N$ on the bifurcation surface with
\begin{equation}
\begin{split}
 \fGamma_N &\equiv \brk{\half \varepsilon_a{}^b \fGamma^a{}_b }_{Bif}\ ,\quad 
 \fR_N \equiv \brk{\half \varepsilon_a{}^b \fR^a{}_b }_{Bif} = d\fGamma_N\, . 
 \end{split}
\end{equation} 

Heuristically,  we can motivate the correction in Eq.~\eqref{eq:WaldTachikawa} from the Chern-Simons terms by thinking of it as descending from a Wald-like formula in one-dimension higher.\footnote{This follows from an identity which holds on the higher-dimensional bifurcation surface :
\begin{equation}
\begin{split}
\left. 2\pi \varepsilon_b{}^a \frac{\partial \fPCFT}{\partial \fR^a{}_b}   \right|_{Bif}
&= \left. d\brk{\  \sum_{k=1}^\infty 8\pi k\  \fGamma_N (d\fGamma_N)^{2k-2} \frac{\partial \fPCFT}{\partial\ \text{tr} \fR^{2k} } } \right|_{Bif}\ . 
 \end{split}
\end{equation} }
In \cite{Tachikawa:2006sz},  Tachikawa outlined an algorithm for modifying Iyer-Wald's derivation\cite{Wald:1993nt,Iyer:1994ys} 
in order to directly derive Eq.~\eqref{eq:WaldTachikawa} along with an explicit derivation in AdS$_3$ case. This algorithm was 
later implemented in higher dimensions by Bonora et al.\cite{Bonora:2011gz}  who however found that it resulted in 
extra non-covariant contributions to Eq.~\eqref{eq:WaldTachikawa} which vanish only in special coordinate systems.\footnote{We note that although the integrand in Eq.~\eqref{eq:WaldTachikawa} is not covariant, its integral over the bifurcation surface is covariant modulo global issues.} This work is motivated by this unsatisfactory state of affairs and to provide a manifestly covariant Noether formalism 
to derive  Eq.~\eqref{eq:WaldTachikawa}.

\subsection*{Summary of results}
In this part, we will summarize our strategy to derive Eq.~\eqref{eq:WaldTachikawa}. We will begin in \S\ref{sec:presymp}
by assigning a covariant pre-symplectic structure over the solutions of Einstein-Maxwell-Chern-Simons equations in 
Eq.~\eqref{eq:EinsteinMaxwellEq}. As we have emphasized before, this is the crucial step in our formalism that 
makes it different from the algorithm proposed by  Tachikawa\cite{Tachikawa:2006sz} which instead works with a 
non-covariant pre-symplectic structure.

In order to write down the pre-symplectic current for Chern-Simons terms, we introduce generalized Hall conductivity
tensors $\{\sHFF,\sHFR,\sHRF,\sHRR\}$ which describe how the Hall currents (defined in Eq.~\eqref{eq:defsHall}) 
vary with field-strengths/curvatures. Let us consider a general variation of the field strengths/curvatures - 
in any dimensions, we can  write the corresponding variation in the Hall currents as
\begin{equation}\label{eq:HallCondIntro}
\begin{split}
\frac{1}{\sqrt{-G}}\, \delta\prn{\sqrt{-G}\ \JH^a } 
&\equiv \half  \prn{\sHFF}^{efa}\cdot  \delta F_{ef}+ 
\half  \prn{\sHFR}_g^{hefa}\ \delta R^g{}_{hef} \, ,\\
\half \frac{1}{\sqrt{-G}}\, \delta\prn{\sqrt{-G}\ (\SpH)^{ab}{}_c }
&\equiv \half  \prn{\sHRF}^{befa}_c \cdot \delta F_{ef}  + 
\half   \prn{\sHRR}_{cg}^{bhefa}\ \delta R^g{}_{hef}\, .  \\
\end{split}
\end{equation}
It follows from the definition of Hall currents in Eq.~\eqref{eq:defsHall} that these Hall conductivities $\{\sHFF,\sHFR,\sHRF,\sHRR\}$  are  completely antisymmetric 
in their last three contravariant  indices (i.e., $efa$ indices in the equations above) : hence, 
they can be thought of as tensor-valued three-forms. Their Hodge-duals are $(2n-2)$-forms in AdS$_{2n+1}$
and they have a simple expression in terms of $\fPCFT$, the anomaly polynomial of the dual CFT :
\begin{equation}\label{eq:fHallCond1}
\begin{split}
 \fsHFF \equiv \frac{\partial^2 \fPCFT}{\partial \fF \partial \fF}\,, & \quad 
\prn{\fsHRR}_{ch}^{bg} \equiv \frac{\partial^2 \fPCFT}{\partial \fR^c{}_b \partial \fR^h{}_g}\ ,  \quad
\prn{\fsHFR}_h^{g} \equiv \prn{\fsHRF}^{g}_h \equiv  
\frac{\partial^2 \fPCFT}{\partial \fF \partial \fR^h{}_g} \, . 
\end{split}
\end{equation}
In terms of these Hall conductivities, we choose a manifestly covariant 
pre-symplectic current corresponding to Chern-Simons terms : 
\begin{equation}\label{eq:PSHAllIntro}
\begin{split}
&(\PSympl^{\ a})_{_H} \\
&\ =
\half \frac{1}{\sqrt{-G}}\, \delta_1 \brk{
\sqrt{-G}\ \prn{\SpH}^{(bc)a} } \delta_2 G_{bc} 
- \half \frac{1}{\sqrt{-G}}\, \delta_2 \brk{
\sqrt{-G}\ \prn{\SpH}^{(bc)a} } \delta_1 G_{bc} \\
&\quad  + \delta_1 A_e \cdot \prn{\sHFF}^{efa}\cdot  \delta_2 A_f
+ \delta_1 \Gamma^c{}_{be} \cdot \prn{\sHRR}^{bhefa}_{cg}\cdot  \delta_2 \Gamma^g{}_{hf} \\ 
&\quad + \delta_1 A_e \cdot \prn{\sHFR}_g^{hefa}\  \delta_2 \Gamma^g{}_{hf}
- \delta_2 A_e \cdot \prn{\sHFR}_g^{hefa}\  \delta_1 \Gamma^g{}_{hf}\, . 
\end{split}
\end{equation}

We will then construct  in \S\ref{sec:diffNoether} a covariant Noether charge consistent with this pre-symplectic current  which takes the form 
\begin{equation}\label{eq:QbarHallIntro}
\begin{split}
(\QNoether^{\ ab})_{H} 
&= \brk{\nabla_h \xi^g \prn{\sHRR}_{gd}^{hcabf} +
\prn{\Lambda+\xi^e A_e} \cdot \prn{\sHFR}_d^{cabf}\ }\delta \Gamma^d{}_{cf}\\
&\quad +  \brk{ \nabla_h \xi^g \prn{\sHRF}_g^{habf}
+\prn{\Lambda+\xi^e A_e} \cdot \prn{\sHFF}^{abf}\ }\cdot \delta A_f \\
&\quad +\half
\brk{ \prn{\SpH}^{(cd)a}\xi^b- \prn{\SpH}^{(cd)b}\xi^a }
\delta G_{cd}  \\
&\quad +\half \frac{\xi^d}{\sqrt{-G}}\,\delta\brk{\sqrt{-G}\  
 G_{cd}\ \prn{\SpH^{acb}+\SpH^{bac}+\SpH^{cab} } } \, ,
\end{split}
\end{equation}
or in terms of differential forms
\begin{equation}\label{eq:QHallExplicitIntro}
\begin{split}
(\fQNoether)_{H} 
&=  \delta \fGamma^c{}_{d} \wedge\brk{\nabla_b \xi^a \frac{\partial^2 \fPCFT}{\partial \fR^a{}_b \partial \fR^c{}_d } +
\prn{\Lambda+ \ic_\xi \fA} \cdot \frac{\partial^2 \fPCFT}{\partial \fF \partial \fR^c{}_d}\ }\\
&\quad + \delta \fA \cdot \brk{ \nabla_b \xi^a \frac{\partial^2 \fPCFT}{ \partial \fR^a{}_b \partial \fF}
+\prn{\Lambda+\ic_\xi \fA} \cdot \frac{\partial^2 \fPCFT}{\partial \fF \partial \fF}\ } \\
&\quad -\half  \delta G_{cd}
\prn{\SpH}^{(cd)a}\ic_\xi \hodge dx_a  \\
&\quad - \xi^d\delta\brk{  
 \half G_{cd}\ \prn{\SpH^{acb}+\SpH^{bac}+\SpH^{cab} }
 \frac{1}{2!} \hodge(dx_a \wedge dx_b) } \, . \\ 
\end{split}
\end{equation}
In particular, we will show in \S\ref{sec:TachikawaDeriv} that this differential Noether 
charge on a bifurcation surface reduces to the Tachikawa formula (the Chern-Simons contribution in Eq.~\eqref{eq:WaldTachikawa}).

\subsection*{Outline}
The organization of the rest of this paper is as follows. In \S\ref{sec:presymp},  we construct 
a pre-symplectic current which leads to a manifestly covariant differential Noether charge 
in \S\ref{sec:diffNoether}.  In particular, we integrate this charge at the bifurcation surface 
to derive the Tachikawa formula for black hole entropy in \S\ref{sec:TachikawaDeriv}. 
In \S\ref{sec:Comparison}, we review the generalization of Lee-Iyer-Wald method 
to Chern-Simons terms as proposed by Tachikawa and compare with our formulation. 
We conclude this paper with some future directions in \S\ref{sec:Conclusion}. 
For reader's convenience, we provide the detail of the derivation of the differential 
Noether charge for Chern-Simons terms in Appendix~\ref{app:QHallDetail}. In Appendix~\ref{sec:Notationsv1} 
we summarize our notation for differential forms and present our formulation in this language.

\section{Pre-symplectic current}\label{sec:presymp}
We will begin our discussion of the Noether procedure which has two main ingredients : the first is a pre-symplectic structure 
on the space of solutions we are interested in and the second being the construction of the Noether charge.  
The main result of this section is the construction of  a covariant pre-symplectic current in the presence of higher dimensional Chern-Simons terms. 

First, in \S\S\ref{ss:basicidea}, we will introduce the idea of a pre-sympletic current.
An explicit example of a pre-sympletic current in the Einstein-Maxwell theory will be given in \S\S\ref{ss:PreSymEM}.
Then we will review in \S\S\ref{ss:LeeWaldPreSym} the discussion of Lee-Iyer-Wald \cite{Lee:1990nz,Iyer:1994ys} in the case of Einstein-Maxwell system while the generalization of the Lee-Iyer-Wald construction to Chern-Simons terms will be presented in \S\ref{sec:Comparison}.
As we will see, however, such a pre-symplectic current in the presence of Chern-Simons terms is non-covariant.
We will thus propose a construction of a manifestly covariant pre-symplectic current
in \S\S\ref{ss:PreSymplHall}.

\subsection{Basic idea}
\label{ss:basicidea}
We start with a dynamical system whose \textbf{equations of motion} we collectively 
represent by $\eom{}$. To be specific, let us consider a theory with dynamical fields
being the metric $G_{ab}$ and a gauge field $A_a$. Then, we can write 
the equations of motion as 
\begin{equation} \label{eq:eqomga}
\begin{split}
 \eom{} = \half\ \delta G_{ab}\ T^{ab} + \delta A_a \cdot J^a\, , 
\end{split}
\end{equation}
where $\{T^{ab},J^a\}$ are some appropriate functionals of the fields $\{ G_{ab},A_a\}$.
The symbol $\slashed{\delta}$ denotes the fact that  $\eom{}$ involves one variation of 
fields. By solutions of these equations of motion, we mean those configurations of  
$\{ G_{ab},A_a\}$ which satisfy $\{T^{ab},J^a\} = 0$. For example, for the Einstein-Maxwell-Chern-Simons system
we are interested in, the equations of motion 
take the form $T^{ab}=(T^{ab})_{\text{Ein-Max}}+\THall^{ab}=0$
and  $J^a=(J^a)_{\text{Ein-Max}}+\JH^a =0$ where
\begin{equation}
\begin{split}
 (T^{ab})_{\text{Ein-Max}}&\equiv-\frac{1}{8\pi\GN}\brk{R^{ab}-\half (R-2\cc)G^{ab}}
+ \frac{1}{\gYM^2}\ \brk{ F^{ac}\cdot F^b{}_c-\quarter G^{ab} F_{cd}\cdot F^{cd} }\, ,\\
 (J^a)_{\text{Ein-Max}}&\equiv- \frac{1}{\gYM^2}\ D_bF^{ab}\ , 
\end{split}
\end{equation}
and the Hall contributions $\THall^{ab}$ and $\JH^a$ are given 
in Eq.~\eqref{eq:defsHall} and just below. 

The next data we will need is the \textbf{pre-symplectic current}\footnote{The adjective
`symplectic' here refers to the fact that this current can be used to define a symplectic
structure on the space of configurations thus allowing us to treat the space of configurations
like a phase-space. The adjective `pre' here refers to the fact that to define the symplectic
structure, often some more work is needed - for example, it is often the case that 
we have to identify the configurations which are gauge equivalent before we can define a 
sensible symplectic structure. We will ignore such complications in the rest of this paper.}
denoted by $(\PSympl)^a$. This is defined such that\footnote{We note that our pre-symplectic
current is negative of the one introduced by Lee-Wald\cite{Lee:1990nz}.} 
\begin{equation}\label{eq:divPS}
\begin{split}
\nabla_a (\PSympl)^a &=
 \frac{1}{\sqrt{-G}}\,\delta_1\prn{\sqrt{-G}\ \eom{2}\ }
 -  \frac{1}{\sqrt{-G}}\,\delta_2\prn{\sqrt{-G}\ \eom{1}\ }\, , 
\end{split}
\end{equation}
i.e., the divergence of the pre-symplectic current is equal to the anti-symmetrized variation 
of the equations of motion. The symbol $\slashed{\delta}^2$ denotes the fact that  $(\PSympl)^a$
involves two variations of  the underlying fields. In our example with Eq.~\eqref{eq:eqomga}, 
this equation becomes
\begin{equation}
\begin{split}
\nabla_a (\PSympl)^a &=
 \half\ \frac{1}{\sqrt{-G}}\,\delta_1\,\brk{\sqrt{-G}\  T^{ab} }\delta_2 G_{ab}\
 -\half\ \frac{1}{\sqrt{-G}}\,\delta_2\,\brk{\sqrt{-G}\  T^{ab} }\delta_1 G_{ab}\\
 &\qquad + \frac{1}{\sqrt{-G}}\,\delta_1\,\brk{\sqrt{-G}\  J^a }\cdot \delta_2 A_a\
 -\frac{1}{\sqrt{-G}}\,\delta_2\,\brk{\sqrt{-G}\  J^a }\cdot \delta_1 A_a \, . \\
\end{split}
\end{equation}

We note that given arbitrary equations of motion, 
the existence of a pre-symplectic current is not always guaranteed. 
However, as we will show below, if, for example,
the equations of motion are derived by varying a manifestly covariant Lagrangian, then
we are guaranteed at least to have a candidate for a pre-symplectic current\cite{Lee:1990nz,Iyer:1994ys}. 

We now proceed towards finding the pre-symplectic current for the system we
are interested in : the Einstein-Maxwell Chern-Simons theory. As a prelude, we will first examine
the simpler case of the Einstein-Maxwell theory.

\subsection{Pre-symplectic current for Einstein-Maxwell theory}
\label{ss:PreSymEM}
The equations of motion for the Einstein-Maxwell theory are given by 
\begin{equation}
\begin{split}
(\eom{})_{\text{Ein-Max}} 
&=  -\half \delta G_{ab}\times \frac{1}{8\pi \GN}   \brk{R^{ab}-\half\prn{R-2\cc}G^{ab}} \\
&\qquad +\half \delta G_{ab}\times \frac{1}{\gYM^2}  
 \brk{ F^{ac}\cdot F^b{}_c -\quarter F^{cd}\cdot F_{cd}\ G^{ab} }\\
&\qquad  -\delta A_a \cdot \frac{1}{\gYM^2} D_b F^{ab} \, . 
\end{split}
\end{equation}

The most commonly used pre-symplectic current for this system is
\begin{equation}\label{eq:PSEinMAx}
\begin{split}
&(\PSympl^{\ a})_{\text{Ein-Max}} =\\ 
&\qquad \frac{1}{\sqrt{-G}}\,\delta_1\prn{\frac{\sqrt{-G}}{8\pi \GN} G^{c[a}\delta^{b]}_d} 
 \ \delta_2 \Gamma^d{}_{cb} 
  -\frac{1}{\sqrt{-G}}\,\delta_2 \prn{\frac{\sqrt{-G}}{8\pi \GN} G^{c[a}\delta^{b]}_d} 
 \ \delta_1 \Gamma^d{}_{cb} \\
&\qquad+\frac{1}{\sqrt{-G}}\,\delta_1 \prn{ \frac{\sqrt{-G}}{\gYM^2}  F^{ab}} \cdot\delta_2  A_b  
-\frac{1}{\sqrt{-G}}\,\delta_2 \prn{ \frac{\sqrt{-G}}{\gYM^2}  F^{ab}} \cdot\delta_1 A_b\,  .  \\
\end{split}
\end{equation}
We will show in the next subsection that this current  obeys Eq.~\eqref{eq:divPS}.

\subsection{Lee-Iyer-Wald prescription : pre-symplectic potential}
\label{ss:LeeWaldPreSym}
It is often convenient to derive the pre-symplectic current from a 
\textbf{pre-symplectic potential} denoted by $\PSymplPot{}^{\ a}$ 
via
\begin{equation}\label{eq:SymplPot}
\begin{split}
 \PSympl^{\ a} &=
 -\frac{1}{\sqrt{-G}}\,\delta_1\brk{\sqrt{-G}\ \PSymplPot{2}^{\ a}\ }
 +  \frac{1}{\sqrt{-G}}\,\delta_2\brk{\sqrt{-G}\ \PSymplPot{1}^{\ a}\ } \, . 
\end{split}
\end{equation}

The existence of such a pre-symplectic potential is closely related to 
the existence of an underlying Lagrangian from which the equations of motion 
can be derived. To see this, we take the divergence of Eq.~\eqref{eq:SymplPot}
so that Eq.~\eqref{eq:divPS} becomes
\begin{equation} 
\frac{1}{\sqrt{-G}}\,\delta_1\brk{\sqrt{-G}\ \prn{\eom{2}+\nabla_a \PSymplPot{2}^{\ a}}\ } 
=  \frac{1}{\sqrt{-G}}\,\delta_2\brk{\sqrt{-G}\ \prn{\eom{1}+\nabla_a \PSymplPot{1}^{\ a}}\ } \, . 
\end{equation}
This is the integrability condition for the existence of a Lagrangian density $\Lag$
such that 
\begin{equation}\label{eq:Lvar}
\begin{split}
 \eom{}+\nabla_a \PSymplPot{}^{\ a}=  \frac{1}{\sqrt{-G}}\,\delta\brk{\sqrt{-G}\ \Lag\ } \, . 
\end{split}
\end{equation}
Thus, the pre-symplectic potential can be thought of as the boundary term that needs to 
be subtracted from the variation of the Lagrangian density to get the equations of motion.
This demonstrates that, for any equations of motion obtained from a Lagrangian, we can 
define a pre-symplectic potential and in turn a pre-symplectic current.

Let us illustrate this with the example of Einstein-Maxwell theory. 
The standard Einstein-Maxwell Lagrangian density  is given by 
\begin{equation}\label{eq:LEinMax}
\begin{split}
\Lag_{_\text{Ein-Max}} &= 
\frac{1}{16\pi \GN}\prn{ R-2\cc }
-\frac{1}{4\gYM^2} F_{ab}\cdot F^{ab} \\
&= - \brk{\half R^d{}_{cab} \frac{G^{c[a}\delta^{b]}_d}{8\pi\GN} + \frac{\cc}{8\pi\GN}
+\quarter  F_{ab} \cdot \frac{F^{ab}}{\gYM^2} } \, . 
\end{split}
\end{equation}
Varying this and adding an appropriate
boundary term give the Einstein-Maxwell equations,viz.,  
\begin{equation}
\begin{split}
\frac{1}{\sqrt{-G}}&\,\delta\,\bigbr{ \sqrt{-G}\ \Lag_{_\text{Ein-Max}}  }
+\nabla_a \left\{ \frac{G^{c[a}\delta^{b]}_d}{8\pi \GN} \ \delta \Gamma^d{}_{cb}
+\frac{F^{ab}}{\gYM^2} \cdot \delta A_b  
\right\}
=(\eom{})_{\text{Ein-Max}}\ .
\end{split}
\end{equation}
We can thus take the pre-symplectic potential as\footnote{It is 
sometimes convenient to write this  pre-symplectic potential as
\begin{equation} \label{eq:symplecticpotentialEM}
\begin{split}
(\PSymplPot{}^{\ a})_{_\text{Ein-Max}} &=  
2\delta \Gamma^d{}_{cb} \frac{\partial \Lag_{_\text{Ein-Max}} }{\partial R^d{}_{cab}} \
+  2\delta A_b  \cdot  \frac{\partial \Lag_{_\text{Ein-Max}} }{\partial F_{ab}}\ , 
\end{split}
\end{equation}
and the corresponding pre-symplectic current as 
\begin{equation}\label{eq:PSEinMAxGen}
\begin{split}
&-(\PSympl^{\ a})_{\text{Ein-Max}} =\\ 
&\qquad\qquad \frac{1}{\sqrt{-G}}\,\delta_1\prn{2\sqrt{-G} \frac{\partial \Lag_{_\text{Ein-Max}} }{\partial R^d{}_{cab}} } 
 \ \delta_2 \Gamma^d{}_{cb} 
  -\frac{1}{\sqrt{-G}}\,\delta_2 \prn{2\sqrt{-G} \frac{\partial \Lag_{_\text{Ein-Max}} }{\partial R^d{}_{cab}} }  
 \ \delta_1 \Gamma^d{}_{cb} \\
&\qquad\qquad+\frac{1}{\sqrt{-G}}\,\delta_1 \prn{ 2\sqrt{-G}  \frac{\partial \Lag_{_\text{Ein-Max}} }{\partial F_{ab}} } \cdot\delta_2  A_b  
-\frac{1}{\sqrt{-G}}\,\delta_2 \prn{ 2\sqrt{-G}  \frac{\partial \Lag_{_\text{Ein-Max}} }{\partial F_{ab}} }  \cdot\delta_1 A_b\,  .  \\
\end{split}
\end{equation}
Written in this form, these Lee-Iyer-Wald pre-symplectic potential and current extend to Lovelock theories.
}
\begin{equation}\label{eq:PSymplPotEinMax}
\begin{split}
-(\PSymplPot{}^{\ a})_{_\text{Ein-Max}} &= \frac{G^{c[a}\delta^{b]}_d}{8\pi \GN} \ \delta \Gamma^d{}_{cb}
+\frac{F^{ab}}{\gYM^2}  \cdot \delta A_b   \, . 
\end{split}
\end{equation}
Varying this potential, we get the pre-symplectic current that we quoted before
in Eq.~\eqref{eq:PSEinMAx}. By construction, this pre-symplectic current then satisfies 
Eq.~\eqref{eq:divPS}.

Given a Lagrangian density, this method then directly gives a candidate for a  pre-symplectic 
current. We note that the pre-symplectic potential computed via an integration by parts
as shown above often depends on the order of integration by parts. A crucial part of the Lee-Iyer-Wald 
prescription\cite{Lee:1990nz,Iyer:1994ys} is, in fact, to prescribe a particular order of integration by parts
which produces covariant pre-symplectic potentials for manifestly covariant Lagrangians.
This, for example, excludes Chern-Simons terms which are of interest to us in this paper. 

\subsection{Pre-symplectic current for Hall currents}\label{ss:PreSymplHall}
Now we want to choose an appropriate pre-symplectic structure for 
the Hall current contribution (i.e, terms in equations of motion coming from varying Chern-Simons terms).
This can be done via the Lee-Iyer-Wald prescription\cite{Lee:1990nz,Iyer:1994ys}
which we had described in our previous subsection. This is the pre-symplectic structure chosen 
by Tachikawa \cite{Tachikawa:2006sz,Bonora:2011gz}. We will compute this  pre-symplectic
current explicitly in  \S\ref{sec:Comparison} and show that  such a  prescription results in a non-covariant 
answer in dimensions greater than three.

We note that a non-covariant pre-symplectic current  is  a serious shortcoming. Usually,  we try to derive the 
symplectic structure on the space of solutions by identifying the directions under which the 
pre-symplectic current is degenerate or non-invertible. With the non-covariant pre-symplectic current, 
this procedure would in general result in a situation  whereby two configurations which are gauge equivalent 
can no more be  identified as a single physical configuration. This breakdown of gauge redundancy 
would then render the theory inconsistent.

In light of these complications, we will adopt  in this subsection  an alternate procedure which produces
a manifestly covariant pre-symplectic current that solves Eq.~\eqref{eq:divPS}. We will refer the reader to  \S\ref{sec:Comparison} for a comparison
of our answer  to the one obtained by Tachikawa's extension of the Lee-Iyer-Wald prescription.

The Hall current contribution to the equations of motion (coming from a variation of  Chern-Simons terms) is given by 
\begin{equation}\label{eq:HallEOM}
\begin{split}
(\eom{})_{_H} 
&=  \half \delta G_{ab}\ (\THall)^{ab} + \delta A_a \cdot \JH^a \\
&= \nabla_a \brk{ \half \delta G_{bc}\prn{\SpH}^{(bc)a} } 
+\half \delta \Gamma^c{}_{ba}\prn{\SpH}^{ab}{}_c + \delta A_a \cdot \JH^a \, .
\end{split}
\end{equation}
In the second equality, we have used 
\begin{equation}\label{eq:THdgIntByParts}
\begin{split}
\half  \delta G_{ab} (\THall)^{ab}
=  \nabla_c\brk{\half  \prn{\SpH}^{(ab)c} \delta G_{ab} }
+ \half  \delta \Gamma^c{}_{ba} \prn{\SpH}^{ab}{}_c \, , 
\end{split}
\end{equation} 
which is obtained from the following relation related to the anti-symmetric property of the spin Hall current 
$\prn{\SpH}^{cab}=- \prn{\SpH}^{cba}$ : 
\begin{equation}
\begin{split}
\delta \Gamma^a{}_{bc} \prn{\SpH}^{cb}{}_a
&
= -  \nabla_a \delta G_{bc}  \prn{\SpH}^{bca}= - (\nabla_c \delta G_{ab})  \prn{\SpH}^{(ab)c} \, . 
\end{split}
\end{equation}

Our strategy for the construction of the pre-symplectic current
is as follows : we begin by computing the anti-symmetrized variation 
of Eq.~\eqref{eq:HallEOM} which should be equal to the divergence of the corresponding
contribution to a pre-symplectic current (see Eq.~\eqref{eq:divPS}) . We use this
fact to write down a manifestly covariant pre-symplectic current which has 
the correct divergence. We first get
\begin{equation} \label{eq:towardoursymplectic}
\begin{split}
&\frac{1}{\sqrt{-G}}\, \delta_1\brk{\sqrt{-G}\ (\eom{2})_{_H}  }
- \frac{1}{\sqrt{-G}}\, \delta_2 \brk{\sqrt{-G}\ (\eom{1})_{_H}  } \\
&\ = \nabla_a \left\{ \half \frac{1}{\sqrt{-G}} \,\delta_1 \brk{
\sqrt{-G}\ \prn{\SpH}^{(bc)a} } \delta_2 G_{bc} 
- \half \frac{1}{\sqrt{-G}}\, \delta_2 \brk{
\sqrt{-G}\ \prn{\SpH}^{(bc)a} } \delta_1 G_{bc}\right\} \\
&\qquad +\half \frac{1}{\sqrt{-G}} \,\delta_1 \brk{
\sqrt{-G}\ \prn{\SpH}^{ab}{}_c } \delta_2 \Gamma^c{}_{ba} 
- \half \frac{1}{\sqrt{-G}}\, \delta_2 \brk{
\sqrt{-G}\ \prn{\SpH}^{ab}{}_c } \delta_1 \Gamma^c{}_{ba} \\
&\qquad + \frac{1}{\sqrt{-G}}\, \delta_1 \brk{
\sqrt{-G}\ \JH^a } \cdot \delta_2 A_a 
- \frac{1}{\sqrt{-G}}\, \delta_2 \brk{
\sqrt{-G}\ \JH^a } \cdot \delta_1 A_a \, . 
\end{split}
\end{equation}
The first line on the right hand side is already in the form of a total divergence. To simplify 
the next two lines, we consider a general variation of the charge and 
spin Hall currents :
\begin{equation}\label{eq:HallCond}
\begin{split}
\frac{1}{\sqrt{-G}} \,\delta\prn{\sqrt{-G}\ \JH^a } 
&\equiv \half  \prn{\sHFF}^{efa}\cdot  \delta F_{ef}+ 
\half  \prn{\sHFR}_g^{hefa}\ \delta R^g{}_{hef} \, ,\\
\half \frac{1}{\sqrt{-G}} \,\delta\prn{\sqrt{-G}\ (\SpH)^{ab}{}_c }
&\equiv \half  \prn{\sHRF}^{befa}_c \cdot \delta F_{ef}  + 
\half   \prn{\sHRR}_{cg}^{bhefa}\ \delta R^g{}_{hef}\, , \\
\end{split}
\end{equation}
where the tensors $\{\sHFF,\sHFR,\sHRF,\sHRR\}$ are 
the generalized Hall conductivities defined in Eq.~\eqref{eq:fHallCond1}. 

Before proceeding, we consider an example to see how these conductivity
tensors look like. Let us take the mixed Chern-Simons term with the anomaly polynomial $\fPCFT= c_{_M} \fF^2\wedge \tr[\fR^2]$ in AdS$_7$ as an example. Then the corresponding charge and spin Hall currents are given by 
\begin{equation}
\begin{split}
\JH^a &= -2 c_{_M} \frac{1}{(2!)^3}\varepsilon^{a\,b_1b_2c_1c_2c_3c_4}
F_{b_1b_2} 
R^e{}_{fc_1c_2}
R^f{}_{e\,c_3c_4} \, , \\
(\SpH)^{ab}{}_c &= -4 c_{_M} \frac{1}{(2!)^3}\varepsilon^{a\,b_1b_2 b_3 b_4c_1c_2}
F_{b_1b_2} F_{b_3 b_4}
R^b{}_{c\,c_1c_2}\, . 
\end{split}
\end{equation}
These expressions can then be varied to give the generalized Hall conductivities
\begin{equation}
\begin{split}
\prn{\sHFF}^{abc} &= -2 c_{_M} \frac{1}{(2!)^2}\varepsilon^{a\,b\,c\,c_1c_2c_3c_4}
R^e{}_{fc_1c_2}
R^f{}_{e\,c_3c_4}\, , \\
\prn{\sHFR}_f^{eabc} &= \prn{\sHRF}_f^{eabc} = 
-4 c_{_M} \frac{1}{(2!)^2}\varepsilon^{a\,b\,c\,b_1b_2c_1c_2}
F_{b_1b_2} R^e{}_{fc_1c_2}\, , \\
\prn{\sHRR}_{fh}^{egabc} &=  
-2 c_{_M} \delta^e_h \delta^g_f\frac{1}{(2!)^2}\varepsilon^{a\,b\,c\,b_1b_2b_3 b_4}
F_{b_1b_2} F_{b_3b_4}\, . 
\end{split}
\end{equation}
Thus, given the Hall currents, it is straightforward to compute the conductivity  
tensors.

A useful property of the conductivity tensors is that their 
covariant divergence (taken with respect to one of its form indices) is zero : 
\begin{equation}\label{eq:divHallCond}
\begin{split}
D_a \prn{\sHFF}^{efa} &= 0\ ,\qquad  
D_a  \prn{\sHFR}_g^{hefa} = 0\ ,\\
D_a  \prn{\sHRF}^{befa}_c  &= 0\ ,\qquad 
D_a  \prn{\sHRR}_{cg}^{bhefa} = 0\ .\
\end{split}
\end{equation}
Further, they satisfy reciprocity type relations
\begin{equation}\label{eq:reciprocity}
\begin{split}
\alpha \cdot \prn{\sHFF}^{efa} \cdot \beta = 
\beta \cdot \prn{\sHFF}^{efa} \cdot \alpha \ , \\
\prn{\sHRR}_{cg}^{bhefa} = \prn{\sHRR}_{gc}^{hbefa}\ ,\qquad 
\prn{\sHFR}^{befa}_c   &=  \prn{\sHRF}^{befa}_c\ .
\end{split}
\end{equation}
Here $\{\alpha,\beta\}$ are two arbitrary scalars transforming in the adjoint representation
of the gauge group. We will need later another set of  identities which are useful 
in getting back  the Hall currents from  the Hall conductivities :
\begin{equation}\label{eq:ixiProp}
\begin{split}
 \delta_f^a \JH^b - \delta_f^b \JH^a
&=  \prn{\sHFF}^{eab}\cdot   F_{ef}+ 
  \prn{\sHFR}_g^{heab}\  R^g{}_{hef} \, , \\
\half \delta_f^a (\SpH)^{bc}{}_d - \half \delta_f^b (\SpH)^{ac}{}_d
&=  \prn{\sHRF}^{ceab}_d \cdot  F_{ef}  + 
 \prn{\sHRR}_{dg}^{cheab}\  R^g{}_{hef} \, . \\
\end{split}
\end{equation}
All these identities can be easily checked for the example of the mixed Chern-Simons term in AdS$_7$.

We now turn to using these properties of the generalized Hall conductivities 
to write down a covariant pre-symplectic current
for arbitrary Chern-Simons terms. The fourth and third lines of Eq.~\eqref{eq:towardoursymplectic} 
are rewritten respectively as 
\begin{equation} \label{eq:4thline}
\begin{split}
\frac{1}{\sqrt{-G}} &\,\delta_1 \brk{
\sqrt{-G}\ \JH^a } \cdot \delta_2 A_a 
- \frac{1}{\sqrt{-G}}\, \delta_2 \brk{
\sqrt{-G}\ \JH^a } \cdot \delta_1 A_a \\
&=  \nabla_a \brk{\delta_1 A_e \cdot \prn{\sHFF}^{aef}\cdot  \delta_2 A_f }\\
&\,\quad + \nabla_a \brk{ \delta_1 A_e \cdot \prn{\sHFR}_g^{haef}\  \delta_2 \Gamma^g{}_{hf}
- \delta_2 A_e \cdot \prn{\sHFR}_g^{haef}\  \delta_1 \Gamma^g{}_{hf} }\\
&\,\quad - \half \delta_2 \Gamma^g{}_{ha}\ \prn{\sHRF}_g^{haef} \cdot  \delta_1 F_{ef} 
+ \half \delta_1 \Gamma^g{}_{ha}\ \prn{\sHRF}_g^{haef} \cdot  \delta_2 F_{ef} \, ,  \\
\end{split}
\end{equation}
and 
\begin{equation} \label{eq:3rdline}
\begin{split}
\half \frac{1}{\sqrt{-G}} &\,\delta_1 \brk{
\sqrt{-G}\ \prn{\SpH}^{ab}{}_c } \delta_2 \Gamma^c{}_{ba} 
- \half \frac{1}{\sqrt{-G}} \,\delta_2 \brk{
\sqrt{-G}\ \prn{\SpH}^{ab}{}_c } \delta_1 \Gamma^c{}_{ba} \\
&=  \nabla_a \brk{
\delta_1 \Gamma^c{}_{be} \cdot \prn{\sHRR}^{bhefa}_{cg}\cdot  \delta_2 \Gamma^g{}_{hf} }\\
&\,\quad + \half \delta_2 \Gamma^g{}_{ha}\ \prn{\sHRF}_g^{haef} \cdot  \delta_1 F_{ef} 
- \half \delta_1 \Gamma^g{}_{ha}\ \prn{\sHRF}_g^{haef} \cdot  \delta_2 F_{ef} \, . \\
\end{split}
\end{equation}

By substituting Eqs.~\eqref{eq:4thline} and \eqref{eq:3rdline}
into Eq.~\eqref{eq:towardoursymplectic}, we finally obtain the pre-symplectic current for our formulation :
\begin{equation}\label{eq:PSHAll}
\begin{split}
&(\PSympl)_{_H}^a \\
&\ =
\half \frac{1}{\sqrt{-G}} \,\delta_1 \brk{
\sqrt{-G}\ \prn{\SpH}^{(bc)a} } \delta_2 G_{bc} 
- \half \frac{1}{\sqrt{-G}}\, \delta_2 \brk{
\sqrt{-G}\ \prn{\SpH}^{(bc)a} } \delta_1 G_{bc} \\
&\,\quad  + \delta_1 A_e \cdot \prn{\sHFF}^{efa}\cdot  \delta_2 A_f
+ \delta_1 \Gamma^c{}_{be} \cdot \prn{\sHRR}^{bhefa}_{cg}\cdot  \delta_2 \Gamma^g{}_{hf} \\ 
&\,\quad + \delta_1 A_e \cdot \prn{\sHFR}_g^{hefa}\  \delta_2 \Gamma^g{}_{hf}
- \delta_2 A_e \cdot \prn{\sHFR}_g^{hefa}\  \delta_1 \Gamma^g{}_{hf}\, . 
\end{split}
\end{equation}
This current is manifestly covariant (recall that variations of the gauge field and Christoffel connection 
transform covariantly) and, by construction, it satisfies
\begin{equation}
\begin{split}
\nabla_a (\PSympl)_{_H}^a &=
 \frac{1}{\sqrt{-G}}\,\delta_1\brk{\sqrt{-G}\ (\eom{2})_{_H} }
 -  \frac{1}{\sqrt{-G}}\,\delta_2\brk{\sqrt{-G}\ (\eom{1})_{_H} }\ \, , 
\end{split}
\end{equation}
as required. Eq.~\eqref{eq:PSHAll} is the main result of this section. In the next section, we will 
use this pre-symplectic current to formulate a manifestly covariant differential Noether charge. 

\section{Noether charge}\label{sec:diffNoether}
Here, we will proceed to the construction of the differential Noether charge for an arbitrary diffeomorphism/gauge transformation.  
After introducing our notations for diffeomorphism/gauge transformation in \S\S\ref{ss:diffeoflav}, we outline the 
basic idea behind the notion of differential Noether charge in \S\S\ref{ss:QNoetherIntro}. As an example, we take up  the 
well-known Lee-Iyer-Wald construction of differential Noether charge for the Einstein-Maxwell system in \S\S\ref{ss:QEinMax}. 
We then turn to sketch the derivation of differential Noether charge for Chern-Simons terms in \S\S\ref{ss:QHall}
relegating most of the details to Appendix~\ref{app:QHallDetail}.

\subsection{Diffeomorphisms and gauge transformations}\label{ss:diffeoflav}
We begin this subsection by introducing our notation for diffeomorphisms and gauge transformations. We will adopt here a notation which admits a natural generalization to non-Abelian gauge transformations.

Given a covariant tensor $\Theta^a{}_b$ transforming in a specific representation
of the gauge group, the action of diffeomorphisms and  gauge transformations is defined via
\begin{equation}\label{eq:deltaTheta}
\begin{split}
\diffF \Theta^a{}_b&=  \lieD_\xi \Theta^a{}_b + [\Theta^a{}_b,\Lambda]\\
& = \xi^c \partial_{c}\Theta^a{}_b
-\prn{\partial_c\xi^a} \Theta^c{}_b+\Theta^a{}_c\, (\partial_b\xi^c)
+ [\Theta^a{}_b,\Lambda] \\
& = \xi^c \prn{\nabla_{c}\Theta^a{}_b+[A_c,\Theta^a{}_b]}
-\prn{\partial_c\xi^a} \Theta^c{}_b+\Theta^a{}_c\,(\partial_b\xi^c)
+ [\Theta^a{}_b,\Lambda+\xi^c A_c]\ .
\end{split}
\end{equation}
Here $\lieD_\xi$ denotes Lie derivative with respect to the vector $\xi^a$ 
parametrizing diffeomorphism, while $\Lambda$ is the gauge parameter in the adjoint representation of the gauge group and the 
commutator $[\Lambda,  \cdot ]$ is the natural adjoint action of the gauge group. We use 
$\chi\equiv \{\xi^a,\Lambda\}$ to jointly denote diffeomorphisms and  gauge transformations.
In the last line of Eq.~\eqref{eq:deltaTheta}, we have defined the covariant derivative
\begin{equation}
\nabla_c \Theta^a{}_b \equiv \partial_{c}\Theta^a{}_b
+ \Gamma^a{}_{dc} \Theta^d{}_b - \Gamma^d{}_{bc} \Theta^a{}_d \, ,
\end{equation}
and the corresponding gauge covariant derivative is
$ D_c \Theta^a{}_b \equiv \nabla_c \Theta^a{}_b + [A_c, \Theta^a{}_b]$. We note 
that the above transformations in Eq.~\eqref{eq:deltaTheta} are covariant provided
$\xi^a$ transforms like a vector and if the combination $\Lambda+ \xi^c A_c$ transforms 
covariantly like a scalar in the adjoint representation.

We have chosen an anti-hermitian basis for the Lie algebra and we have suppressed 
all gauge indices for convenience. We can write $\Lambda = -i T_A (\Lambda_{Real})^A$ etc. 
with  $T_A$ being the standard hermitian gauge group generators. Further, a trace over gauge indices
is indicated by  ``$\cdot$'', a `center dot'. For example, if $A_a \equiv A_a^C (-iT_C)$ 
and $J^a \equiv J^{a C} (iT_C)$, then $ A_a \cdot J^a =  A_a^C  J^{a D} \ \text{Tr}\prn{T_CT_D}$.
In this anti-hermitian convention, Abelian  gauge fields are purely imaginary, i.e., 
if $A_a$ is an Abelian gauge field, then $A_a= -i(A_a)_{Real}$. The corresponding 
Abelian current would be $J^a= i(J^a)_{Real}$ so that $ A_a \cdot J^a =  (A_a)_{Real} (J^a)_{Real}$.
The Abelian action on a field $\Psi_q$ with Abelian charge $q$ is 
given by $[\Lambda, \Psi_q] = q \Lambda \Psi_q = -iq(\Lambda)_{Real} \Psi_q$.

We now turn to the action of $\diffF$ on various quantities. 
We can write the transformation of background gauge field/metric as
\begin{equation}\label{eq:deltaAg}
\begin{split}
\diffF A_{a}&=  \lieD_\xi A_a + [A_{a},\Lambda]+\partial_{a} \Lambda= \partial_{a} \Lambda + [A_{a},\Lambda]+ A_{c}\partial_{a}
\xi^{c} + \xi^{c} \partial_{c} A_{a}\\
&= \partial_{a} \prn{\Lambda +\xi^{c} A_{c} }+ [A_{a},\ \Lambda +\xi^{c} A_{c}]+\xi^c F_{ca}\,,\\
\diffF G_{ab} &= \lieD_\xi G_{ab}= G_{cb}\partial_a \xi^c+ G_{ac}\partial_b \xi^c + \xi^c \partial_c G_{ab}\\
&=\nabla_{a} \xi_{b} + \nabla_{b} \xi_{a} \, ,\\
\end{split}
\end{equation}
where $F_{ab}$ denotes the field strength for $A_a$. 

The Christoffel connection $\Gamma^b{}_{ac}$,
being the unique torsionless and metric-compatible connection, is
completely determined by the background metric $G_{ac}$ as 
\begin{equation}\label{eq:GammaMet}
\begin{split}
\Gamma^b{}_{ac} \equiv \half  G^{bd}\brk{ \partial_a G_{cd} + \partial_c G_{ad} - \partial_d G_{ac} } \, .
\end{split}
\end{equation}
We will use this connection and the associated covariant derivative from now on unless specified.
For the Christoffel connection, the transformation is
\begin{equation}\label{eq:deltaGamma}
\begin{split}
\diffF \Gamma^a{}_{bc}
&=\half  G^{ad} \brk{\nabla_b \prn{\nabla_{c} \xi_{d} + \nabla_{d} \xi_{c}} 
+\nabla_c\prn{\nabla_{b} \xi_{d} + \nabla_d \xi_{b} } 
- \nabla_d\prn{\nabla_{b} \xi_{c} + \nabla_{c} \xi_{b}} }
\\
&=\nabla_c\nabla_b \xi^a + \xi^d R^a{}_{bdc}\, .\\
\end{split}
\end{equation}
The field strength $F_{ab}$ and the curvature tensor $R^d{}_{abc}$ transform as  covariant tensors under
gauge transformations and diffeomorphisms. We also note that variations of gauge field and Christoffel 
connection, i.e. $\delta A_a$ and $\delta\Gamma^a{}_{bc}$, transform covariantly. 

It is sometimes convenient to focus only on the non-covariant part of transformations and drop the
covariant parts. We denote this part by defining
\begin{equation}
\begin{split}
\delnc(\ldots) \equiv \diffF (\ldots) - \lieD_\xi (\ldots) -[\ldots,\, \Lambda] \, .
\end{split}
\end{equation}
It follows from Eq.~\eqref{eq:deltaTheta} that $\delnc\Theta^a{}_b =0 $ for
any covariant tensor field $\Theta^a{}_b$. The connections, on the other hand, transform
non-covariantly as
\begin{equation}
\begin{split}
\delnc A_a
&=\partial_a\Lambda\, ,  \qquad 
\delnc \Gamma^a{}_{bc}
=\partial_c\partial_b \xi^a\, .\\
\end{split}
\end{equation}

\subsection{Differential Noether charge}\label{ss:QNoetherIntro}

We begin with Eq.~\eqref{eq:divPS} where we take the second variation to be
the diffeomorphism/gauge variation $\diffF$ generated by $\chi=\{\xi^a,\Lambda\}$ :
\begin{equation}\label{eq:divOmegaChi}
\begin{split}
\nabla_a (\PSymChi)^a&=
 \frac{1}{\sqrt{-G}}\,\delta \prn{\sqrt{-G}\ \eomChi\ }
 -  \frac{1}{\sqrt{-G}}\,\diffF\prn{\sqrt{-G}\ \eom{}\ } \, . 
\end{split}
\end{equation}
This implies that $\nabla_a (\PSymChi)^a \simeq 0$ on-shell, i.e., once we impose the 
equations of motion $\eom{}=0$ . Here we have used the symbol $\simeq$
to denote that the equality holds only on-shell.

Assuming that there are no  cohomological obstructions, the statement 
$\nabla_a (\PSymChi)^a \simeq 0$
implies that there exists a $(\QNoether)^{ab}$ such that 
\begin{equation}\label{eq:QDefOn-Shell}
\begin{split}
-\nabla_b\ (\QNoether)^{ab} \simeq (\PSymChi)^a
\end{split}
\end{equation}
with $(\QNoether)^{ab}=-(\QNoether)^{ba}$. We will call a $(\QNoether)^{ab}$ 
that satisfies the above equation  as the \textbf{differential Noether charge} 
associated with the diffeomorphism/gauge variation $\diffF$. Our aim is to 
construct the differential Noether charge for the Einstein-Maxwell-Chern-Simons
system so that one can use it to assign charges to solutions of this system.

Often in the construction of the differential Noether charge, it is convenient
to work off-shell and impose the equations of motion $\eom{}=0$ only at the end.
In order to do this, we need to lift Eq.~\eqref{eq:QDefOn-Shell} to an 
off-shell statement. In case of covariant equations of motion, this can be done 
using Noether's theorem.

Assuming $\eom{}$ transforms as a scalar under diffeomorphism/gauge transformations, the 
second term on the right hand side of Eq.~\eqref{eq:divOmegaChi} becomes
\begin{equation} 
\begin{split}
\frac{1}{\sqrt{-G}}\,\diffF\prn{\sqrt{-G}\ \eom{}\ }
= (\nabla_a \xi^a) \eom{} + \xi^a\nabla_a (\eom{})
= \nabla_a\prn{ \xi^a\ \eom{}\ } \, , 
\end{split}
\end{equation}
so that we can write Eq.~\eqref{eq:divOmegaChi} as
\begin{equation}\label{eq:divOmegaChi1}
\begin{split}
\nabla_a\brk{\ (\PSymChi)^a+ \xi^a\ \eom{}\ } =  \frac{1}{\sqrt{-G}}\,\delta \prn{\sqrt{-G}\ \eomChi\ } \, . 
\end{split}
\end{equation}

We then turn our attention to the right hand side of Eq.~\eqref{eq:divOmegaChi1}. To simplify this 
term we now invoke  Noether's theorem (see \cite{Compere:2007az} for example). Noether's theorem asserts the following:
\textbf{
If the system under 
question is invariant under the diffeomorphism/gauge variation $\diffF$ generated 
by $\chi=\{\xi^a,\Lambda\}$,\footnote{The reader should note that theories with Chern-Simons terms are included in this set.}
 then there exists a Noether current $\N^a$ such that 
$\eomChi = \nabla_a \N^a$. Further, we can always choose an on-shell-vanishing $\N^a$, i.e.,
a Noether current $\N^a$ such that $\N^a\simeq 0$.}

To illustrate this, we consider the example where the only dynamical fields are metric and  
gauge fields. We then have
\begin{equation}
\begin{split}
\eomChi &= \half T^{ab}\diffF G_{ab} + J^a\cdot \diffF A_a \\
&= T^{ab} \nabla_b \xi_a + \xi_a J_b\cdot F^{ab} +J^a\cdot D_a \prn{ \Lambda+\xi^c A_c}
+\half T^{ab} \prn{\nabla_a \xi_b- \nabla_b \xi_a} \\
&= \nabla_a \brk{ \xi_b T^{ba} + J^a\cdot \prn{ \Lambda+\xi^c A_c} }\\
&\qquad -\xi_a \brk{\nabla_b T^{ab}-J_b\cdot F^{ab}} - (D_a J^a)\cdot \prn{ \Lambda+\xi^c A_c}\\
&\qquad +\half \prn{T^{ab}-T^{ba}}\nabla_a \xi_b \, .\\
\end{split}
\end{equation}
If the equations of motion describe a system which is diffeomorphism/gauge  invariant, then we 
have the following Bianchi identities (which hold off-shell) :
\begin{equation}
\begin{split}
\nabla_b T^{ab}=J_b\cdot F^{ab}\ , \qquad  T^{ab}=T^{ba} \ , \qquad D_a J^a =0\,.
\end{split}
\end{equation}
We can therefore choose the on-shell-vanishing Noether current as
\begin{equation}
\begin{split}
\N^a=\xi_b T^{ab} + J^a\cdot \prn{ \Lambda+\xi^c A_c} \, . 
\end{split}
\end{equation}

Let us now use such a Noether current to simplify Eq.~\eqref{eq:divOmegaChi1} to
\begin{equation}
\begin{split}
\nabla_a \brk{\ (\PSymChi)^a+ \xi^a\ \eom{}\ }
 = \nabla_a\brk{ \frac{1}{\sqrt{-G}}\delta \prn{\sqrt{-G}\  \N^a} } \, ,
\end{split}
\end{equation}
or
\begin{equation}
\begin{split}
\nabla_a\brk{\ (\PSymChi)^a+ \xi^a\ \eom{}\
-\frac{1}{\sqrt{-G}}\delta \prn{\sqrt{-G}\  \N^a } }  = 0\,.
\end{split}
\end{equation}
This is the off-shell generalization 
of the statement $\nabla_a (\PSymChi)^a \simeq 0$. The statement
$-\nabla_b\ (\QNoether)^{ab} \simeq (\PSymChi)^a$ then generalizes off-shell
to 
\begin{equation} \label{eq:definingeqofnoethercharge}
\begin{split}
-\nabla_b\ (\QNoether)^{ab} = (\PSymChi)^a+ \xi^a\ \eom{}\
-\frac{1}{\sqrt{-G}}\,\delta \prn{\sqrt{-G}\  \N^a } \, , 
\end{split}
\end{equation}
or
\begin{equation}
\begin{split}
\frac{1}{\sqrt{-G}}\,\delta \prn{\sqrt{-G}\  \N^a }
= (\PSymChi)^a+ \xi^a\ \eom{}\
+\nabla_b\ (\QNoether)^{ab} \, . 
\end{split}
\end{equation}
This expression shows that $ (\QNoether)^{ab}$ can be thought of as the surface contribution to the variation of the 
Noether current $\N^a$, thus justifying the name `differential Noether charge'. 
In the following subsections, we will apply the above Noether procedure to the Einstein-Maxwell-Chern-Simons system.

\subsection{Differential Noether charge for Einstein-Maxwell system}\label{ss:QEinMax}
Our goal in this subsection is to compute the differential Noether charge for Einstein-Maxwell system.
We begin by writing down the on-shell vanishing Noether current for this system :
\begin{equation}
\begin{split}
(\N^a)_{\text{Ein-Max}} &= -\frac{\xi^b}{8\pi \GN}\brk{ R^a{}_b -\half \prn{ R-2\cc } \delta^a_b } 
\\
&\qquad +\frac{\xi^b}{\gYM^2}\brk{ F^{ac}\cdot F_{bc} -\quarter F^{cd}\cdot F_{cd} \delta^a_b }
-\frac{\prn{\Lambda+\xi^c A_c}}{\gYM^2}\cdot D_b F^{ab} \, . 
\end{split}
\end{equation}
This current is, by construction, proportional to the Einstein-Maxwell equations of motion.
Hence, it vanishes on any solution of Einstein-Maxwell system (thus the adjective `on-shell
vanishing').

Let us rewrite this Noether current in a suggestive way :
\begin{equation} \label{eq:noethereinmax}
\begin{split}
(\N^a)_{\text{Ein-Max}} &= \xi^a \brk{ \frac{1}{16\pi \GN}\prn{ R-2\cc } -\frac{1}{4\gYM^2} F^{cd}\cdot F_{cd} }  \\
&\qquad-\frac{1}{8\pi \GN} \xi^b R^a{}_b 
+\frac{1}{\gYM^2} \brk{ \xi^b F^{ac}\cdot F_{bc} -\prn{\Lambda+\xi^c A_c}\cdot D_b F^{ab} } \, .
\end{split}
\end{equation}
We recognize the standard Einstein-Maxwell Lagrangian density (see Eq.~\eqref{eq:LEinMax}) in the first line of 
Eq.~\eqref{eq:noethereinmax}.
On the other hand we can rewrite the second line of Eq.~\eqref{eq:noethereinmax} using the following identities~:
\begin{equation}
\begin{split}
 F^{ab} \cdot\diffF A_b  &= F^{ab} \cdot D_b\prn{\Lambda+\xi^c A_c} +  F^{ac} \cdot \xi^b F_{bc} \\
&= \nabla_b \brk{ \prn{\Lambda+\xi^c A_c}\cdot  F^{ab} } + \xi^b  F^{ac} \cdot F_{bc}  
-\prn{\Lambda+\xi^c A_c}\cdot D_b F^{ab} \, , \\
\end{split}
\end{equation}
and 
\begin{equation}
\begin{split}
G^{c[a}\delta^{b]}_d\ \diffF \Gamma^d{}_{cb} &=G^{c[a}\delta^{b]}_d\ \nabla_b \nabla_c \xi^d
+ G^{c[a}\delta^{b]}_d\ \xi^f R^d{}_{cfb}\\
&=  \nabla_b \prn{ G^{c[a}\delta^{b]}_d\ \nabla_c \xi^d} -\xi^b R^a{}_b \, , 
\end{split}
\end{equation}
so that we have the following expression for the Noether current $(\N^a)_{\text{Ein-Max}} $  : 
\begin{equation}
\begin{split}
(\N^a)_{\text{Ein-Max}} &= \xi^a \Lag_{\text{Ein-Max}}+\frac{1}{8\pi \GN} G^{c[a}\delta^{b]}_d\ \diffF \Gamma^d{}_{cb} 
+\frac{1}{\gYM^2}  F^{ab} \cdot\diffF A_b \\
&\qquad-\nabla_b \bigbr{ \frac{1}{8\pi \GN} G^{c[a}\delta^{b]}_d\ \nabla_c \xi^d
+\frac{1}{\gYM^2} F^{ab} \cdot \prn{\Lambda+\xi^c A_c}  
}\, . 
\end{split}
\end{equation}

We recognize that the pre-symplectic potential for the Einstein-Maxwell system 
(see Eq.~\eqref{eq:PSymplPotEinMax} with the variation set equal to a diffeomorphism/gauge 
variation) is 
\begin{equation}
\begin{split}
-(\PSymplPotChi^{\ a})_{\text{Ein-Max}} &=\frac{1}{8\pi \GN} G^{c[a}\delta^{b]}_d\ \diffF \Gamma^d{}_{cb} 
+\frac{1}{\gYM^2}  F^{ab} \cdot\diffF A_b  \, . 
\end{split}
\end{equation}
Further, defining 
\begin{equation} \label{eq:komarpartEM}
\begin{split}
-(\Komar^{\ ab})_{\text{Ein-Max}} &\equiv \frac{1}{8\pi \GN} G^{c[a}\delta^{b]}_d\ \nabla_c \xi^d
+\frac{1}{\gYM^2} F^{ab} \cdot \prn{\Lambda+\xi^c A_c}  \, , 
\end{split}
\end{equation}
which is often termed the \textbf{Komar charge}, we can write the Einstein-Maxwell contribution 
to the Noether current in the following form :
\begin{equation}\label{eq:KomarForm}
\begin{split}
 (\N^a)_{\text{Ein-Max}} &= \bigbr{\xi^a \Lag - (\PSymplPotChi)^a +\nabla_b \Komar^{\ ab} }_{\text{Ein-Max}} .
\end{split}
\end{equation}
We will call this form of decomposition for the on-shell-vanishing Noether current
as the \textbf{Komar decomposition}. 

The Komar decomposition exists for any covariant Lagrangian.\footnote{
For example, if we rewrite the Einstein-Maxwell Komar charge as
\begin{equation}
\begin{split}
(\Komar^{\ ab})_{\text{Ein-Max}} &\equiv 2 \frac{\partial \Lag_{_\text{Ein-Max}} }{\partial R^d{}_{cab}} \ \nabla_c \xi^d
+2\frac{\partial \Lag_{_\text{Ein-Max}} }{\partial F_{ab}}\ \cdot \prn{\Lambda+\xi^c A_c}  \, , 
\end{split}
\end{equation}
then this form can be extended to Lovelock theories.
}
To see why this might be the case, we consider the divergence of the vector $\xi^a \Lag - (\PSymplPotChi)^a$ :
\begin{equation}
\begin{split}
 \nabla_a \brk{ \xi^a \Lag - (\PSymplPotChi)^a }
 = \nabla_a \prn{ \xi^a \Lag } -\frac{1}{\sqrt{-G}}\,\diffF \prn{ \sqrt{-G}\ \Lag }+\eomChi
 =\eomChi \, , 
\end{split}
\end{equation}
where we have used the fact that, if $\Lag$ is a scalar, then 
$(\sqrt{-G})^{-1}\,\diffF \prn{ \sqrt{-G}\ \Lag } =\nabla_a \prn{ \xi^a \Lag } $.
This shows that the vector $\xi^a \Lag - (\PSymplPotChi)^a$ is a Noether current by itself
(we note however that it is not on-shell vanishing). The Komar decomposition then follows from 
the statement that any two Noether currents differ by a total divergence. 

The Komar decomposition plays an important role in the Lee-Iyer-Wald method for computing 
differential Noether charge. We consider a general variation applied to the Komar
decomposition written in the form
\begin{equation}\label{eq:komardecomposition}
\begin{split}
-\nabla_b \Komar^{\ ab}&= \xi^a \Lag - (\PSymplPotChi)^a - \N^a .
\end{split}
\end{equation}
Then we get 
\begin{equation}\label{eq:simplifykomar}
\begin{split}
-\nabla_b &\brk{\frac{1}{\sqrt{-G}}\,\delta\prn{ \sqrt{-G}\ \Komar^{\ ab} } }\\
&=\xi^a\ \frac{1}{\sqrt{-G}}\delta\prn{ \sqrt{-G}\ \Lag} 
-\frac{1}{\sqrt{-G}}\,\delta\brk{ \sqrt{-G}\ (\PSymplPotChi)^a}  - \frac{1}{\sqrt{-G}}\,\delta\prn{ \sqrt{-G}\ \N^a } .
\end{split}
\end{equation}
We now use Eq.~\eqref{eq:Lvar} as well as the  
following relation to rewrite the first and second terms on the right hand side : 
\begin{equation}
\begin{split}
&\frac{1}{\sqrt{-G}}\,\delta\brk{ \sqrt{-G}\ (\PSymplPotChi)^a}\\
&\ = \frac{1}{\sqrt{-G}}\,\diffF\brk{ \sqrt{-G}\ (\PSymplPot{})^a}-(\PSymChi)^a\\
&\ = (\PSymplPot{})^a\nabla_b \xi^b+ \xi^b \nabla_b (\PSymplPot{})^a 
-(\nabla_b \xi^a)(\PSymplPot{})^b -(\PSymChi)^a\\
&\ = - \nabla_b\brk{ \xi^a (\PSymplPot{})^b - (\PSymplPot{})^a \xi^b  }+
\xi^a \nabla_b (\PSymplPot{})^b-(\PSymChi)^a \, . \\
\end{split}
\end{equation}
Here we have assumed $(\PSymplPotChi)^a$ transforms like a vector and is invariant under
gauge transformations. Substituting these relations back into Eq.~\eqref{eq:simplifykomar}, we get 
\begin{equation}
\begin{split}
-\nabla_b &\brk{\frac{1}{\sqrt{-G}}\,\delta\prn{ \sqrt{-G}\ \Komar^{\ ab} } + 
\xi^a (\PSymplPot{})^b - (\PSymplPot{})^a \xi^b }\\
&\quad=(\PSymChi)^a + \xi^a\ \eom{}  - \frac{1}{\sqrt{-G}}\delta\prn{ \sqrt{-G}\ \N^a } .
\end{split}
\end{equation}
From this expression, we can then identify the differential Noether charge according to the Lee-Iyer-Wald prescription
(for systems with covariant Lagrangian and symplectic potential) as
\begin{equation} \label{eq:noetherintermsofkomar}
\begin{split}
\QNoether^{\ ab} = \frac{1}{\sqrt{-G}}\,\delta\prn{ \sqrt{-G}\ \Komar^{\ ab} } 
+ \xi^a (\PSymplPot{})^b - (\PSymplPot{})^a \xi^b .
\end{split}
\end{equation}

For the Einstein-Maxwell system, by using Eqs.~\eqref{eq:PSymplPotEinMax} and \eqref{eq:komarpartEM}, 
this differential Noether charge is written as 
\begin{equation}\label{eq:QbarEinMax}
\begin{split}
(&\QNoether^{\ ab})_{_\text{Ein-Max}} \\
&\ =
-\frac{1}{\sqrt{-G}}\,\delta\brk{ \sqrt{-G}\ \prn{ \frac{1}{8\pi \GN} G^{c[a}\delta^{b]}_d\ \nabla_c \xi^d
+\frac{1}{\gYM^2} F^{ab} \cdot \prn{\Lambda+\xi^c A_c}  }}\\
&\qquad-\xi^a\brk{ \frac{1}{8\pi \GN} G^{c[b}\delta^{f]}_d\ \delta \Gamma^d{}_{cf}
+\frac{1}{\gYM^2} F^{bf} \cdot \delta A_f  }\\
&\qquad +\xi^b\brk{ \frac{1}{8\pi \GN} G^{c[a}\delta^{f]}_d\ \delta \Gamma^d{}_{cf}
+\frac{1}{\gYM^2} F^{af} \cdot \delta A_f  } \, . 
\end{split}
\end{equation}

\subsection{Differential Noether charge for Chern-Simons terms}
\label{ss:QHall}
A differential Noether charge for theories with Chern-Simons terms was constructed by Tachikawa by generalizing the Lee-Iyer-Wald method\cite{Tachikawa:2006sz,Bonora:2011gz}. As we will demonstrate in \S\ref{sec:Comparison}, this charge however turns out to be non-covariant.  In this subsection, we will
instead  construct  a manifestly covariant differential Noether charge.

We now proceed to evaluate the contribution to the differential Noether charge from Chern-Simons terms 
by directly using its relation with the pre-symplectic current :
\begin{equation} \label{eq:definingeqnoethercs}
\begin{split}
-\nabla_b (\QNoether^{\ ab})_{H} 
=(\PSymChi)_{_H}^a + \xi^a (\eom{})_{_H}-\frac{1}{\sqrt{-G}}\delta\brk{\sqrt{-G}\ \N_{_H}^a} \, . 
\end{split}
\end{equation}

The Hall contribution $\N_{_H}^a$ to the on-shell vanishing Noether current for this system is given by 
\begin{equation}\label{eq:NoetherHall}
\begin{split}
\N_{_H}^a
&=  \xi_b (\THall)^{ab} + \prn{\Lambda+\xi^c A_c} \cdot \JH^a 
\\
&= \nabla_b \brk{ \half \xi_c\prn{\SpH^{acb}+\SpH^{bac}+\SpH^{cab} } } 
+\half  \SpH^{(bc)a} \diffF G_{bc}  \\
&\qquad +\half \nabla_c \xi^d \prn{\SpH}^{ac}{}_d + \prn{\Lambda+\xi^c A_c} \cdot \JH^a \, . \\
\end{split}
\end{equation}
Using this along with our covariant expression for the pre-symplectic current, we get
\begin{equation} \label{eq:righthandsideofdifnoetherhall}
\begin{split}
&(\PSymChi)_{_H}^a + \xi^a (\eom{})_{_H}-\frac{1}{\sqrt{-G}}\,\delta\brk{\sqrt{-G}\ \N_H^a}\\
&= -\nabla_b\Bigl\{\ \half
\brk{  \prn{\SpH}^{(cd)a}\xi^b -\xi^a \prn{\SpH}^{(cd)b} }
\delta G_{cd} \Bigr.\\
&\qquad \qquad \Bigl. +\half \frac{\xi^d}{\sqrt{-G}}\delta\brk{\sqrt{-G}\  
 G_{cd}\ \prn{\SpH^{acb}+\SpH^{bac}+\SpH^{cab} } }\ \Bigr\}\\
&\qquad + \xi^a \brk{\half\delta \Gamma^d{}_{cb} \prn{\SpH}^{bc}{}_d +\delta A_b \cdot \JH^b }  \\
&\qquad -\frac{1}{\sqrt{-G}}\delta\brk{\sqrt{-G}\ 
\prn{\half \nabla_c \xi^d \prn{\SpH}^{ac}{}_d + \prn{\Lambda+\xi^c A_c} \cdot \JH^a}\ } \\
&\qquad  + \delta A_e \cdot \prn{\sHFF}^{efa}\cdot  \diffF A_f
+ \delta \Gamma^c{}_{be} \cdot \prn{\sHRR}^{bhefa}_{cg}\cdot  \diffF \Gamma^g{}_{hf} \\ 
&\qquad + \delta A_e \cdot \prn{\sHFR}_g^{hefa}\  \diffF \Gamma^g{}_{hf}
- \diffF A_e \cdot \prn{\sHFR}_g^{hefa}\  \delta \Gamma^g{}_{hf} \, . 
\end{split}
\end{equation}
The details of the computation that lead to this expression can be found in Appendix~\ref{app:QHallDetail}.

We can then express the right hand side of Eq.~\eqref{eq:righthandsideofdifnoetherhall} 
as a total divergence to give
\begin{equation}\label{eq:QbarHall}
\begin{split}
(\QNoether^{\ ab})_{H} 
&= \brk{\nabla_h \xi^g \prn{\sHRR}_{gd}^{hcabf} +
\prn{\Lambda+\xi^e A_e} \cdot \prn{\sHFR}_d^{cabf}\ }\delta \Gamma^d{}_{cf}\\
&\quad +  \brk{ \nabla_h \xi^g \prn{\sHRF}_g^{habf}
+\prn{\Lambda+\xi^e A_e} \cdot \prn{\sHFF}^{abf}\ }\cdot \delta A_f \\
&\quad +\half
\brk{ \prn{\SpH}^{(cd)a}\xi^b- \prn{\SpH}^{(cd)b}\xi^a }
\delta G_{cd}  \\
&\quad +\half \frac{\xi^d}{\sqrt{-G}}\delta\brk{\sqrt{-G}\  
 G_{cd}\ \prn{\SpH^{acb}+\SpH^{bac}+\SpH^{cab} } } \, ,
\end{split}
\end{equation}
which is a manifestly covariant differential Noether charge as required.
In Appendix~\ref{sec:Notationsv1}, we convert the above expression into differential forms :
\begin{equation}\label{eq:QHallExplicit}
\begin{split}
(\fQNoether)_{H} 
&=  \delta \fGamma^c{}_{d} \wedge\brk{\nabla_b \xi^a \frac{\partial^2 \fPCFT}{\partial \fR^a{}_b \partial \fR^c{}_d } +
\prn{\Lambda+ \ic_\xi \fA} \cdot \frac{\partial^2 \fPCFT}{\partial \fF \partial \fR^c{}_d}\ }\\
&\quad + \delta \fA \cdot \brk{ \nabla_b \xi^a \frac{\partial^2 \fPCFT}{ \partial \fR^a{}_b \partial \fF}
+\prn{\Lambda+\ic_\xi \fA} \cdot \frac{\partial^2 \fPCFT}{\partial \fF \partial \fF}\ } \\
&\quad -\half  \delta G_{cd}
\prn{\SpH}^{(cd)a}\ic_\xi \hodge dx_a  \\
&\quad - \xi^d\delta\brk{  
 \half G_{cd}\ \prn{\SpH^{acb}+\SpH^{bac}+\SpH^{cab} }
 \frac{1}{2!} \hodge(dx_a \wedge dx_b) } \, . \\ 
\end{split}
\end{equation}
This manifestly covariant differential Noether charge is the main result of this paper. 

In the next section, we evaluate this differential Noether charge at the bifurcation surface to derive the Tachikawa formula (see Eq.~\eqref{eq:WaldTachikawa}). In an accompanying paper\cite{Azeyanagi:2014}, we will use Eq.~\eqref{eq:QHallExplicit} to covariantly assign both entropy and charges to the black hole solutions found in \cite{Azeyanagi:2013xea} and compare them against the dual CFT expectations.

\section{A covariant derivation of Tachikawa formula}
\label{sec:TachikawaDeriv}
In this section,  we give a covariant derivation of the Tachikawa formula described 
in Eq.~\eqref{eq:WaldTachikawa}  using our differential Noether charge Eq.~\eqref{eq:QHallExplicit}. Our derivation here is aimed at 
neatly  sidestepping various issues raised by Bonora et al.\cite{Bonora:2011gz}  regarding Tachikawa's extension 
of the Lee-Iyer-Wald method. In particular,  unlike the derivation in \cite{Bonora:2011gz}, 
we do not have to pass to special  coordinates/gauges in order to suppress various non-covariant contributions that arise in Tachikawa's proposal.

Let us begin by recalling that at the bifurcation surface we have $\xi^a |_{Bif}=0$ and $(\Lambda+\xi^a A_a) |_{Bif}=\LambdaH+ \xiH^a A_a =0$\,.  Thus Eq.~\eqref{eq:QHallExplicit} reduces to 
\begin{equation}\label{eq:QHallBif}
\begin{split}
\left. (\fQNoether)_{H} \right |_{Bif}
&=  -2\pi \varepsilon^a{}_b \brk{ \delta \fGamma^c{}_{d}  \frac{\partial^2 \fPCFT}{\partial \fR^a{}_b \partial \fR^c{}_d } 
+ \delta \fA \cdot \frac{\partial^2 \fPCFT}{ \partial \fR^a{}_b \partial \fF} } \, . 
\end{split}
\end{equation} Here we have used
\begin{equation}
\nabla_b \xi^a |_{Bif}= 2 \pi \varepsilon_b{}^a=-2\pi \varepsilon^a{}_b \, ,
\end{equation} where $\varepsilon$ is the binormal to the bifurcation surface.
Furthermore, following  \cite{Wald:1993nt,Iyer:1994ys},  we have
$\delta \varepsilon^a{}_b= 0$, since $\xi^a |_{Bif}=0$ while $\delta \xi^a=0$ everywhere.

For simplicity, let us first start with the single trace case where
$\fP_{CFT}=c_{_M}\, \fF^l \wedge \tr[\fR^{2k}]$ in AdS$_{2n+1}$ with $n=2k+l-1$.
Derivatives of the anomaly polynomial with respect to the curvature two-form and 
the $U(1)$ field strength are given respectively by 
\begin{equation} \label{eq:ddPdRdR}
\begin{split}
\frac{\partial^2 \fP_{CFT}}{ \partial \fR^a{}_b \partial \fF} &= c_{_M}\,(2 k l ) \fF^{l-1}\wedge (\fR^{2k-1})^b{}_a\, , \\
\frac{\partial^2 \fP_{CFT}}{\partial \fR^a{}_b\partial \fR^c{}_d} &=c_{_M}\,( 2  k) \fF^l\wedge
\sum_{m=0}^{2k-2} (\fR^{m})^b{}_c (\fR^{2k-2-m})^d{}_a\, , 
\end{split}
\end{equation}
where we take $(\fR^0)^b{}_c\equiv \delta^b{}_c$.
Substituting the above into Eq.~\eqref{eq:QHallBif} yields\footnote{In the following, the binormal $\varepsilon$ inside the traces should be interpreted as the matrix  $\varepsilon^a{}_b$.}
\begin{equation} \label{eqQHallBifsimplified}
\begin{split} 
\left. (\fQNoether)_{H} \right |_{Bif}
&=  -2\pi  c_{_M}\,(2k)  \left\{\fF^l\wedge
\sum_{m=0}^{2k-2} \tr[ \delta \fGamma \fR^{2k-2-m}\varepsilon \fR^{m} ] \Bigr. \Bigl.+\ l\ \delta \fA \cdot  \fF^{l-1}\wedge \tr[ \varepsilon \fR^{2k-1} ] \right\} \, .\\
\end{split}
\end{equation}

Let us now discuss the pull-backs of $\fGamma$ and $\fR$ onto the bifurcation surface. Since $\nabla_c \varepsilon_{ab}=0$ at the bifurcation surface, the induced metrics on the tangent and normal bundle are also covariantly constant.
Therefore, the restriction of the covariant derivative onto the tangent (normal) bundle is equal to the covariant derivative constructed out of the tangent (normal) bundle metric. This implies that at the bifurcation surface $\varepsilon^a{}_b \fR^b{}_c=\fR^{a}{}_b \varepsilon^{b}{}_c=  -\varepsilon^a{}_b \varepsilon^b{}_c \fR_N$ where $-2 \fR_N\equiv\tr(\varepsilon \fR) |_{Bif}$.
The normal bundle curvature  $\fR_N$ satisfies $\fR_N=d\fGamma_N$ where $-2\fGamma_N\equiv \tr\prn{ \varepsilon \fGamma    } |_{Bif}$.\footnote{
To show $\fR_N=d\fGamma_N$, we can use the decomposition of the binormal $\varepsilon_{ab}=\rho_a \xiH_b - \xiH_a \rho_b$ for vectors $\rho$ and $\xiH$ satisfying $\rho_a\rho^a=\xiH_a \xiH^a=0$ and $\rho^a \xiH_a= 1$ at the bifurcation surface. Then using an equivalent definition of $\fGamma_N\equiv \rho^b \nabla_c \xiH_b dx^c$, one can show that $\fR_N=d\fGamma_N$.
For more details on the properties of the normal bundle at the bifurcation surface, see  \cite{Booth:2006bn}.
}
We will exploit below this factorization between the normal and tangent bundle at the bifurcation surface.

Using this, at the bifurcation  we have $\fR^{2k-2-m}\varepsilon \fR^{m} = \fR_N^{2k-2} \varepsilon$ and $ \tr\prn{ \varepsilon \fR^{2k-1} }  =-2 \fR_N^{2k-1}$.
These allow us to rewrite Eq.~\eqref{eqQHallBifsimplified} in the following form :
\begin{equation}
\begin{split}\label{eq:tachikawaBifFormula}
\left. (\fQNoether)_{H} \right |_{Bif}
&=  2\pi  c_{_M}\,(2k) \fR_N^{2k-2}  \wedge \Bigl\{-(2k-1) \fF^l\wedge  
 \tr[ \delta \fGamma   \varepsilon ]+\ 2 l\ \delta \fA \cdot  \fF^{l-1}\wedge  \fR_N  \Bigr\} \\
&=  8\pi k\  c_{_M}\, \fR_N^{2k-2}  \wedge \Bigl\{(2k-1) \fF^l\wedge  
  \delta \fGamma_N   +\ l\ \delta \fA \cdot  \fF^{l-1}\wedge  d\fGamma_N  \Bigr\} \\ 
&=  8\pi k\  c_{_M}\, \fR_N^{2k-2}  \wedge \Bigl\{(2k-1) \fF^l\wedge  
  \delta \fGamma_N   +\ l\ \delta \fF \cdot  \fF^{l-1}\wedge  \fGamma_N  \Bigr\}  \\
  &\qquad +8\pi k\  c_{_M}\, d\brk{ l\  \delta \fA \cdot  \fF^{l-1}\wedge  \fGamma_N  \fR_N^{2k-2}  
  } \\   
&= \delta\brk{  \ 8\pi k\ \fGamma_N   \fR_N^{2k-2}  \wedge \  c_{_M}\ \fF^l } \\
  &\qquad + 8\pi k\  c_{_M}\, d\Bigl\{(2k-2)  \fF^l\wedge \fGamma_N \fR_N^{2k-3}  
  \delta \fGamma_N+ l\  \delta \fA \cdot  \fF^{l-1}\wedge  \fGamma_N  \fR_N^{2k-2}  
  \Bigr\} \\    
 &= \delta\brk{  8\pi k\ \fGamma_N   \fR_N^{2k-2}  \wedge \  \frac{\partial \fP_{CFT}}{ \partial\ \tr \fR^{2k}}} +d(\ldots)  \, ,
\end{split} 
\end{equation} which agrees with the result of Tachikawa in  \cite{Tachikawa:2006sz}. 

We now use induction to generalize this formula to the case with multiple traces. First, 
we denote the anomaly polynomial as $\fPCFT = \tilde{\fP} \wedge \tr[\fR^{2k_0}]$
and assume that $\tilde{\fP}$ contributes to the black hole entropy via the Tachikawa formula. 
For example, for the anomaly polynomial  $\fP_{CFT}=c_{_M} \,\fF^l \wedge\tr[\fR^{2k_1}]\wedge \tr[\fR^{2k_0}]$, 
the term $\tilde{\fP}$ is given by  $\tilde{\fP}=c_{_M}\,\fF^l\wedge \tr[\fR^{2k_1}]$. 
Now we will show that 
$\fPCFT$ also contributes to the entropy via the Tachikawa formula as in the last line of Eq.~\eqref{eq:tachikawaBifFormula}. In this case, Eq.~\eqref{eq:QHallBif} becomes
\begin{equation}
\begin{split}\label{eq:multitracedeltaQbif}
\left. (\fQNoether)_{H} \right|_{Bif}
 &= \brk{\  \delta\prn{  \sum_{k=1}^\infty 8\pi k\ \fGamma_N   \fR_N^{2k-2}  \wedge \  \frac{\partial \tilde{\fP} }{ \partial\ \tr \fR^{2k}}} +d(\ldots)   }  
 \wedge  \tr[\fR^{2k_0}]\\
 &+  \tilde{\fP} \wedge \brk{\  \delta\prn{  8\pi k_0\ \fGamma_N   \fR_N^{2k_0-2}  } +d(\ldots)   } \\
 & -  (2k_0) \left\{   \tr\left[\delta\fGamma \frac{\partial \tilde{\fP}}{\partial \fR}\right] 2\pi \wedge \tr\left[\varepsilon \fR^{2k_0-1}\right] +  \tr\left[\delta\fGamma \fR^{2k_0-1}\right] \wedge 2\pi \tr\left[\varepsilon   \frac{\partial \tilde{\fP}}{\partial \fR}\right] \right\} .\\
\end{split} 
\end{equation}
The first line above correspond to the terms where both of the derivatives on the anomaly polynomial (with respect to the curvature two-form) act on 
$\tilde{\fP} $
 while the second line corresponds to the terms where both derivatives act on $\tr[\fR^{2k_0}]$. The two terms in the third line account for the cases where one derivative acts on $\tilde{\fP}$ while the other on $\tr[\fR^{2k_0}]$.
 
As a next step, we use
\begin{equation}
\begin{split}
-2\pi (2k_0)\wedge \tr[\varepsilon \fR^{2k_0-1}]  &= 8\pi k_0\   \fR_N^{2k_0-1} \, , \\
-2\pi\ \tr\left[\varepsilon   \frac{\partial \tilde{\fP}}{\partial \fR}\right] 
 &= \sum_{k=1}^\infty 8\pi k\   \fR_N^{2k-1}  \wedge \  \frac{\partial \tilde{\fP} }{ \partial\ \tr \fR^{2k}} \, , 
 \end{split}
\end{equation}
to write the last line of Eq.~\eqref{eq:multitracedeltaQbif} as 
\begin{equation}
\begin{split}
&   \tr\left[\delta\fGamma \frac{\partial \tilde{\fP}}{\partial \fR}\right] 8\pi k_0\   \fR_N^{2k_0-1} +  (2k_0) \tr\left[\delta\fGamma \fR^{2k_0-1}\right] \wedge  \sum_{k=1}^\infty 8\pi k\   \fR_N^{2k-1}  \wedge \  \frac{\partial \tilde{\fP} }{ \partial\ \tr \fR^{2k}}    \\
  &\quad =   \tr\left[\delta\fR \frac{\partial \tilde{\fP}}{\partial \fR}\right] 8\pi k_0\   \fGamma_N \fR_N^{2k_0-2} +  (2k_0) \tr[\delta\fR \fR^{2k_0-1}] \wedge  \sum_{k=1}^\infty 8\pi k\   \fR_N^{2k-1}  \wedge \  \frac{\partial \tilde{\fP} }{ \partial\ \tr \fR^{2k}}    \\
&\quad\,\,   
  - d\left\{   \tr\left[\delta\fGamma \frac{\partial \tilde{\fP}}{\partial \fR}\right] 8\pi k_0\  \fGamma_N \fR_N^{2k_0-2} +  (2k_0) \tr[\delta\fGamma \fR^{2k_0-1}] \wedge  \sum_{k=1}^\infty 8\pi k\   \fGamma_N \fR_N^{2k-2}  \wedge \  \frac{\partial \tilde{\fP} }{ \partial\ \tr \fR^{2k}}   \right\} \\
 &\quad =  \delta \tilde{\fP} \wedge 8\pi k_0\   \fGamma_N \fR_N^{2k_0-2} +\delta \tr[ \fR^{2k_0}] \wedge  \sum_{k=1}^\infty 8\pi k\   \fR_N^{2k-1}  \wedge \  \frac{\partial \tilde{\fP} }{ \partial\ \tr \fR^{2k}}  +d(\ldots)  \, .
 \end{split}
\end{equation}
Finally, substituting the above expression into Eq.~\eqref{eq:multitracedeltaQbif}, we obtain 
\begin{equation}
\begin{split}
\left. (\fQNoether)_{H} \right|_{Bif}
 &= \delta\prn{  \sum_{k=1}^\infty 8\pi k\ \fGamma_N   \fR_N^{2k-2}  \wedge \  \frac{\partial \tilde{\fP} }{ \partial\ \tr \fR^{2k}}}  
 \wedge  \tr[\fR^{2k_0}]\\
 &\quad+  \tilde{\fP} \wedge   \delta\prn{  8\pi k_0\ \fGamma_N   \fR_N^{2k_0-2}  }  \\
 &\quad  +\delta \tilde{\fP} \wedge 8\pi k_0\   \fGamma_N \fR_N^{2k_0-2} +\delta \tr[ \fR^{2k_0}] \wedge  \sum_{k=1}^\infty 8\pi k\   \fR_N^{2k-1}  \wedge \  \frac{\partial \tilde{\fP} }{ \partial\ \tr \fR^{2k}} +d(\ldots) \\
 &= \delta\prn{  \sum_{k=1}^\infty 8\pi k\ \fGamma_N   \fR_N^{2k-2}  \wedge \  \frac{\partial \fPCFT }{ \partial\ \tr \fR^{2k}}}  
  +d(\ldots)\, ,  \\
\end{split}
\end{equation}
which proves the Tachikawa formula for Chern-Simons contribution to entropy.

%%%%%%%%%%%%%%%%%%%%%%%

\section{Tachikawa's extension of Lee-Iyer-Wald method : a comparison}
\label{sec:Comparison}

Now, we will review the generalization of the Lee-Iyer-Wald method to Chern-Simons
terms as proposed by Tachikawa\cite{Tachikawa:2006sz}. The reader should also consult
\cite{Bonora:2011gz} where a detailed exposition of this method is given. Since the 
discussion below is somewhat long and technical, we will begin by 
summarizing what we do in this section.

\subsection{Summary of this section}
The primary aim of this section is to take our discussion about the formulation of Noether charge
for theories with Chern-Simons terms and link it with the previous proposals in the literature -
mainly the references \cite{Tachikawa:2006sz,Bonora:2011gz}. We begin with an 
explicit implementation of Tachikawa's prescription for the most general 
Chern-Simons term. Our analysis can be thought of as a straightforward 
generalization of the analysis in \cite{Bonora:2011gz}.

We will show that our Noether charge agrees with the Noether charge of \cite{Tachikawa:2006sz} 
in AdS$_3$ where Tachikawa's extension has been primarily applied. However, 
Tachikawa's extension for the formulation of Noether charge
deviates from our method in higher dimensions
by various additional non-covariant contributions which we will explicitly compute below. 
Thus, our prescription
neatly resolves this non-covariance issue with higher dimensional Chern-Simons terms that was noted 
by the authors of \cite{Bonora:2011gz}. 

We will now present two main analytical results of this section that lead to the 
conclusions above. The first is the relation between our covariant pre-symplectic current $(\fPSympl)_{_H}$ and the 
non-covariant pre-symplectic current $(\fPSympl)^{IWT}_{_H}$ from Tachikawa's extension:
\begin{equation} \label{eq:IWTpresymplecticcurrent}
\begin{split}
(\fPSympl)^{IWT}_{_H}
&\ = (\fPSympl)_{_H} \\
&\,\quad+ d\bigbr{ \delta_1 \fA \cdot \frac{\partial^2\ICS}{\partial \fF \partial \fF} \cdot  \delta_2 \fA
+ \delta_1 \fGamma^c{}_b \frac{\partial^2\ICS}{\partial \fR^c{}_b \partial \fR^g{}_h}
  \delta_2 \fGamma^g{}_h }\\ 
&\,\quad + d\bigbr{ \delta_1 \fA \cdot \frac{\partial^2\ICS}{\partial \fF \partial \fR^g{}_h}\  \delta_2 \fGamma^g{}_h
- \delta_2 \fA \cdot \frac{\partial^2\ICS}{\partial \fF \partial \fR^g{}_h}\  \delta_1 \fGamma^g{}_h}\, . 
\end{split}
\end{equation} 
Since the derivatives of the Chern-Simons action $\ICS$ in Eq.~\eqref{eq:IWTpresymplecticcurrent}
contain $\fA$ or $\fGamma^a{}_b$, 
this expression shows that $(\fPSympl)^{IWT}_{_H}$ is non-covariant in AdS$_5$ and higher\footnote{We note that the non-covariant contributions vanish for AdS$_3$.} and 
that non-covariance enters as a boundary contribution. 

The second result we derive is the relation between our covariant differential Noether
charge $(\fQNoether)_{_H}$ and the non-covariant differential Noether
charge $(\fQNoether)^{IWT}_{_H}$ from Tachikawa's extension :
\begin{equation}
\begin{split}
(\fQNoether)^{IWT}_{_H} 
&= (\fQNoether)_{_H}  \\
&\,\quad -\bigbr{ \delta \fA \cdot \frac{\partial^2\ICS}{\partial \fF \partial \fF} \cdot  \diffF \fA
+ \delta \fGamma^c{}_b \frac{\partial^2\ICS}{\partial \fR^c{}_b \partial \fR^g{}_h}
  \diffF \fGamma^g{}_h }\\ 
&\,\quad -\bigbr{ \delta \fA \cdot \frac{\partial^2\ICS}{\partial \fF \partial \fR^g{}_h}\  \diffF \fGamma^g{}_h
- \diffF \fA \cdot \frac{\partial^2\ICS}{\partial \fF \partial \fR^g{}_h}\  \delta \fGamma^g{}_h}\\
&\,\quad+d\bigbr{\fZ
+\delta \fA \cdot \frac{\partial^2\ICS}{\partial \fF \partial \fF} \cdot (\Lambda+\ic_\xi\fA)
+ \delta \fGamma^c{}_b \frac{\partial^2\ICS}{\partial \fR^c{}_b \partial \fR^g{}_h}
  \nabla_h \xi^g }\\ 
&\,\quad +d\bigbr{
\delta \fA \cdot \frac{\partial^2\ICS}{\partial \fF \partial \fR^g{}_h}\  \nabla_h \xi^g
+ \delta \fGamma^g{}_h  \frac{\partial^2\ICS}{\partial \fR^g{}_h \partial \fF}
\cdot (\Lambda+\ic_\xi\fA)  }\, .\\
\end{split}
\end{equation}
This expression shows that $(\fQNoether)^{IWT}_{_H}$ is non-covariant in AdS$_5$ and higher and 
that non-covariance enters both as a bulk and a boundary contribution. The boundary contribution
is however ambiguous in Tachikawa's extension which is represented by an arbitrary term $\fZ$
in the expression above. 

In the rest of this section, we will derive these analytical results. Since the non-covariance
of Tachikawa's extension complicates the formulation of Noether charge for general Chern-Simons terms, 
we will work entirely with differential forms throughout this section. 

\subsection{Pre-symplectic current in Tachikawa's extension}
As a first step of the comparison, we compute the deviation of the pre-symplectic currents 
constructed by Tachikawa's extension from ours.

For the Chern-Simons terms, the Lagrangian form is given by $\fLag_{_H}=\ICS\,$.
The corresponding equations of motion form are given by converting Eq.~\eqref{eq:HallEOM} into 
differential forms~:
\begin{equation}\label{eq:fHallEOM}
\begin{split}
(\feom{})_{_H}
&= -(\eom{})_{_H}\ \hodge 1\\ 
&= - d\brk{\half \SpH^{(ab)c}\delta G_{ab}\ \hodge dx_c }
-\half \delta \fGamma^a{}_b\prn{\hodge\fSpH}^b{}_a - \delta \fA \cdot \hodge\fJH \, . 
\end{split}
\end{equation}
The pre-symplectic potential $\fPSymplPot{}$ is given as
\begin{equation}\label{eq:fPSymplPotH}
\begin{split}
\fPSymplPot{} &= \half \SpH^{(ab)c}\delta G_{ab}\ \hodge dx_c 
+ \delta \fGamma^a{}_b \frac{\partial \ICS}{ \partial \fR^a{}_b}
+ \delta \fA \cdot \frac{\partial \ICS}{ \partial \fF} \, . 
\end{split}
\end{equation}
Following Tachikawa\cite{Tachikawa:2006sz}, we then define the 
pre-symplectic current as
\begin{equation}\label{eq:IWTPSHAllF}
\begin{split}
(\fPSympl)^{IWT}_{_H}
&\equiv -\delta_1 (\fPSymplPot{2})_{_H} + \delta_2 (\fPSymplPot{1})_{_H}\\
&=
-\half \delta_1 \brk{
 \prn{\SpH}^{(bc)a} \ \hodge dx_a } \delta_2 G_{bc} 
+ \half  \delta_2 \brk{
\prn{\SpH}^{(bc)a}\ \hodge dx_a } \delta_1 G_{bc} \\
&\,\quad  + \delta_1 \fA \cdot \fsHFF \cdot  \delta_2 \fA
+ \delta_1 \fGamma^c{}_b \cdot \prn{\fsHRR}^{bh}_{cg}\cdot  \delta_2 \fGamma^g{}_h \\ 
&\,\quad + \delta_1 \fA \cdot \prn{\fsHFR}_g^{h}\  \delta_2 \fGamma^g{}_h
- \delta_2 \fA \cdot \prn{\fsHFR}_g^h\  \delta_1 \fGamma^g{}_h\\
&\,\quad + d\bigbr{ \delta_1 \fA \cdot \frac{\partial^2\ICS}{\partial \fF \partial \fF} \cdot  \delta_2 \fA
+ \delta_1 \fGamma^c{}_b \frac{\partial^2\ICS}{\partial \fR^c{}_b \partial \fR^g{}_h}
  \delta_2 \fGamma^g{}_h }\\ 
&\,\quad + d\bigbr{ \delta_1 \fA \cdot \frac{\partial^2\ICS}{\partial \fF \partial \fR^g{}_h}\  \delta_2 \fGamma^g{}_h
- \delta_2 \fA \cdot \frac{\partial^2\ICS}{\partial \fF \partial \fR^g{}_h}\  \delta_1 \fGamma^g{}_h}\, , 
\end{split}
\end{equation}
where we have used the following identities to simplify the expression : 
\begin{equation}\label{eq:CondICSIds}
\begin{split}
\fsHFF&=\frac{\partial^2 \fPCFT}{ \partial \fF  \partial \fF}
= \frac{\partial^2 \ICS}{ \partial \fA  \partial \fF}
+ \frac{\partial^2 \ICS}{ \partial \fF  \partial \fA}
+ D\prn{ \frac{\partial^2 \ICS}{ \partial \fF \partial \fF} } \, , \\
(\fsHRF)^b_a &=\frac{\partial^2 \fPCFT}{ \partial \fR^a{}_b  \partial \fF} 
= \frac{\partial^2 \ICS}{ \partial \fGamma^a{}_b  \partial \fF}
+ \frac{\partial^2 \ICS}{ \partial \fR^a{}_b   \partial \fA}
+ D\prn{ \frac{\partial^2 \ICS}{ \partial \fR^a{}_b  \partial \fF} } \, , \\
(\fsHFR)^d_c&=\frac{\partial^2 \fPCFT}{\partial \fF \partial \fR^c{}_d  } 
= \frac{\partial^2 \ICS}{\partial \fA \partial \fR^c{}_d   }
+ \frac{\partial^2 \ICS}{\partial \fF \partial \fGamma^c{}_d  }
+ D\prn{ \frac{\partial^2 \ICS}{   \partial \fF \partial \fR^c{}_d} } \, , \\
(\fsHRR)^{bd}_{ac}&=\frac{\partial^2 \fPCFT}{\partial \fR^a{}_b \partial \fR^c{}_d  } 
= \frac{\partial^2 \ICS}{\partial \fGamma^a{}_b \partial \fR^c{}_d   }
+ \frac{\partial^2 \ICS}{\partial \fR^a{}_b\partial \fGamma^c{}_d  }
+ D\prn{ \frac{\partial^2 \ICS}{  \partial \fR^a{}_b \partial \fR^c{}_d} } \, .\\
\end{split}
\end{equation}
Comparing against the pre-symplectic current $(\fPSympl)_{_H}$ derived in 
Eq.~\eqref{eq:PSHAllF}, the relation between these two pre-symplectic currents is given by 
\begin{equation}
\begin{split}
(\fPSympl)^{IWT}_{_H}
&\ = (\fPSympl)_{_H}
+ d\bigbr{ \delta_1 \fA \cdot \frac{\partial^2\ICS}{\partial \fF \partial \fF} \cdot  \delta_2 \fA
+ \delta_1 \fGamma^c{}_b \frac{\partial^2\ICS}{\partial \fR^c{}_b \partial \fR^g{}_h}
  \delta_2 \fGamma^g{}_h }\\ 
&\qquad + d\bigbr{ \delta_1 \fA \cdot \frac{\partial^2\ICS}{\partial \fF \partial \fR^g{}_h}\  \delta_2 \fGamma^g{}_h
- \delta_2 \fA \cdot \frac{\partial^2\ICS}{\partial \fF \partial \fR^g{}_h}\  \delta_1 \fGamma^g{}_h} \, . 
\end{split}
\end{equation}
Unlike $(\fPSympl)_{_H}$, the current $(\fPSympl)^{IWT}_{_H}$ is not covariant under gauge
and diffeomorphisms in AdS$_{d+1}$ for $d\geq 4$. Since it is 
covariant up to a boundary contribution, the pre-symplectic
structure defined via its integral would be invariant. 
As we will see below, however, these boundary contributions do contribute to the Noether charge
thus affecting the covariance of $\fQNoether$. Discarding by hand this 
non-covariant boundary contribution in $(\fPSympl)^{IWT}_{_H}$, we 
get back $(\fPSympl)_{_H}$.

\subsection{Komar decomposition for Chern-Simons terms}
We next move on to the Komar decomposition. Following \cite{Tachikawa:2006sz}
we begin by constructing two differential forms $\fXi$ and $\fSigma$
defined via\footnote{We note that our $\fSigma$ is negative of the one used in
in \cite{Tachikawa:2006sz,Bonora:2011gz}.}
\begin{equation}\label{eq:fXiSigmaDef}
\begin{split}
\delnc \ICS &= d\fXi  \, , \\
\delta \fXi   &= \delnc (\fPSymplPot{})_{_H}  + d \fSigma\ \, .
\end{split}
\end{equation}
A direct computation gives
\begin{equation}\label{eq:fXiSigma}
\begin{split}
\fXi &\equiv  \Lambda  \cdot \frac{\partial \ICS}{ \partial \fA}
  + \partial_b \xi^a\ \frac{\partial \ICS}{ \partial \fGamma^a{}_b} - d\fY \, ,  \\
\fSigma &\equiv
\delta \fA \cdot \brk{
 \frac{\partial^2 \ICS}{\partial \fF \partial \fA} \cdot \Lambda
 + \frac{\partial^2 \ICS}{\partial \fF \partial \fGamma^c{}_d}\ \partial_d \xi^c }  \\
  & \qquad + \delta \fGamma^a{}_b\  \brk{ \frac{\partial^2 \ICS}{\partial \fR^a{}_b \partial \fA} \cdot \Lambda
 +  \frac{\partial^2 \ICS}{\partial \fR^a{}_b \partial \fGamma^c{}_d}\ \partial_d \xi^c } -\delta \fY+ d\fZ \, , \\
\end{split}
\end{equation}
where $\fY$ and $\fZ$ are arbitrary forms undetermined by this procedure. We note that 
$\fXi$ encodes the consistent anomaly of the dual CFT and thus we will refer to $\fXi$ 
as the consistent anomaly form. 

Using these forms, we can write
\begin{equation}
\begin{split}
-d\hodge\fN_{_H}=(\feom{\chi})_{_H} = \diffF \ICS - d(\fPSymplPot{\chi})_{_H} = d\brk{\ic_\xi \ICS +\fXi-(\fPSymplPot{\chi})_{_H} }\, . 
\end{split}
\end{equation}
The Komar decomposition takes the form
\begin{equation}\label{eq:fKomarDecompH}
\begin{split}
-\hodge\fN_{_H}= \ic_\xi \ICS +\fXi-(\fPSymplPot{\chi})_{_H} +  d\fKomar \, . 
\end{split}
\end{equation}
Here $\fN_{_H}$ is the Chern-Simons part of the on-shell vanishing Noether current 
(see Eq.~\eqref{eq:NoetherHall}) given by
\begin{equation}\label{eq:fNoetherHall}
\begin{split}
\hodge\fN_{_H} &= d \brk{ \half \xi_c\prn{\SpH^{acb}+\SpH^{bac}+\SpH^{cab} }
\frac{1}{2!} \hodge(dx_a \wedge dx_b) } 
+\half  \SpH^{(bc)a} \diffF G_{bc}\ \hodge dx_a  \\
&\qquad +\half \nabla_c \xi^d \prn{\hodge\fSpH}^c{}_d + \prn{\Lambda+\ic_\xi \fA} \cdot\ \hodge\fJH \, . 
\end{split}
\end{equation}
This gives the Komar charge as
\begin{equation}\label{eq:fKomarHall}
\begin{split}
(\fKomar)_{_H} &\equiv \fY
+ \nabla_b \xi^a\ \frac{\partial \ICS}{ \partial \fR^a{}_b} 
+  (\Lambda + \ic_\xi \fA)  \cdot \frac{\partial \ICS}{ \partial \fF}\\
&\ -\half \xi_c\prn{\SpH^{acb}+\SpH^{bac}+\SpH^{cab} }
\frac{1}{2!} \hodge(dx_a \wedge dx_b) \, .
\end{split}
\end{equation}

We note that the Komar term in this case is completely ambiguous by an addition of
an arbitrary form $\fY$. Further, we remind the reader that, as emphasized by  Bonora et al.\cite{Bonora:2011gz},
this expression does not directly lead to the analogue of Wald entropy, unless the form $\fY$ 
is suitably chosen  and one works in a special set of coordinates/gauges. 
More explicitly, this can be done in a two-step process : 
\begin{enumerate}
\item First,  fix various ambiguities in Tachikawa's extension (the objects $\fY$ and $\fZ$ above)
so that the forms $\fXi$ and $\fSigma$ are proportional to $d\Lambda$ and $d(\partial_a\xi^b)$.
\item Next, pass to a certain special gauges/coordinate systems where $d\Lambda=0$ and $d(\partial_a\xi^b)=0$ at the bifurcation surface, so that the forms $\fXi$ and $\fSigma$ 
in the non-covariant Tachikawa's extension vanish.
\end{enumerate}
Once the forms $\fXi$ and $\fSigma$ are made to vanish by these two steps, one can derive an 
effective Komar charge for Chern-Simons terms from which one can derive the Tachikawa formula for Chern Simons contribution to entropy in this special set of gauges/co-ordinates \cite{Bonora:2011gz}.

\subsection{Differential Noether charge for Chern-Simons terms in Tachikawa's extension}
Finally, we evaluate the difference between the differential Noether charges 
constructed by the two methods.

We begin with the Komar decomposition for the Chern-Simons term Eq.~\eqref{eq:fKomarDecompH}
which we rewrite as
\begin{equation}
\begin{split}
-d(\fKomar)_{_H}= \ic_\xi \ICS +\fXi-(\fPSymplPot{\chi})_{_H} + \hodge\fN_{_H}  \, . 
\end{split}
\end{equation}
Now we consider the variation of this expression. By using 
\begin{equation}
\begin{split}
\delta&\brk{\fXi-(\fPSymplPot{\chi})_{_H}}- d\fSigma\\
&\qquad=  \delnc(\fPSymplPot{})_{_H}-\diffF(\fPSymplPot{})_{_H}+(\fPSymChi)^{IWT}_{_H}\\
&\qquad= (\fPSymChi)^{IWT}_{_H} - \ic_\xi d(\fPSymplPot{})_{_H} -  d\ic_\xi(\fPSymplPot{})_{_H}\, , 
\end{split}
\end{equation}
and
\begin{equation}
\begin{split}
\ic_\xi \delta\ICS +   d\ic_\xi(\fPSymplPot{})_{_H}
&= \ic_\xi (\feom{})_{_H} + \ic_\xi d(\fPSymplPot{})_{_H} +  d\ic_\xi(\fPSymplPot{})_{_H} \, , 
\end{split}
\end{equation}
we have the following expression : 
\begin{equation}
\begin{split}
-d\brk{\delta\fKomar+\fSigma-\ic_\xi \fPSymplPot{}}_{_H}
&= (\fPSymChi)^{IWT}_{_H}+\ic_\xi (\feom{})_{_H} +  \delta\hodge\fN_{_H} \, .
\end{split}
\end{equation}
Thus, we obtain the differential Noether charge according to Tachikawa's prescription 
as
\begin{equation}\label{eq:fQIWT}
\begin{split}
(\fQNoether)^{IWT}_{_H} &= \delta(\fKomar)_{_H}+\fSigma-\ic_\xi \prn{\fPSymplPot{}}_{_H}\, . 
\end{split}
\end{equation}
Using Eqs.~\eqref{eq:fPSymplPotH}, \eqref{eq:fXiSigma} and \eqref{eq:fKomarHall}, this simplifies to
\begin{equation}
\begin{split}
(\fQNoether)^{IWT}_{_H} 
&= \delta \fGamma^d{}_{c} \wedge\brk{\nabla_h \xi^g \prn{\fsHRR}_{gd}^{hc} +
\prn{\Lambda+ \ic_\xi \fA} \cdot \prn{\fsHFR}_d^{c}\ }\\
&\,\quad + \delta \fA \cdot \brk{ \nabla_h \xi^g \prn{\fsHRF}_g^{h}
+\prn{\Lambda+\ic_\xi \fA} \cdot \prn{\fsHFF}\ } \\
&\,\quad -\half  \delta G_{cd}
\prn{\SpH}^{(cd)a}\ic_\xi \hodge dx_a  \\
&\,\quad - \xi^d\delta\brk{  
 \half G_{cd}\ \prn{\SpH^{acb}+\SpH^{bac}+\SpH^{cab} }
 \frac{1}{2!} \hodge(dx_a \wedge dx_b) }  \\
&\,\quad-\bigbr{ \delta \fA \cdot \frac{\partial^2\ICS}{\partial \fF \partial \fF} \cdot  \diffF \fA
+ \delta \fGamma^c{}_b \frac{\partial^2\ICS}{\partial \fR^c{}_b \partial \fR^g{}_h}
  \diffF \fGamma^g{}_h }\\ 
&\,\quad -\bigbr{ \delta \fA \cdot \frac{\partial^2\ICS}{\partial \fF \partial \fR^g{}_h}\  \diffF \fGamma^g{}_h
- \diffF \fA \cdot \frac{\partial^2\ICS}{\partial \fF \partial \fR^g{}_h}\  \delta \fGamma^g{}_h}\\
&\,\quad+d\bigbr{\fZ
+\delta \fA \cdot \frac{\partial^2\ICS}{\partial \fF \partial \fF} \cdot (\Lambda+\ic_\xi\fA)
+ \delta \fGamma^c{}_b \frac{\partial^2\ICS}{\partial \fR^c{}_b \partial \fR^g{}_h}
  \nabla_h \xi^g }\\ 
&\,\quad +d\bigbr{
\delta \fA \cdot \frac{\partial^2\ICS}{\partial \fF \partial \fR^g{}_h}\  \nabla_h \xi^g
+ \delta \fGamma^g{}_h  \frac{\partial^2\ICS}{\partial \fR^g{}_h \partial \fF}
\cdot (\Lambda+\ic_\xi\fA)  } \, . \\
\end{split}
\end{equation}
Here we have also used the identities for the generalized Hall conductivities summarized in Eq.~\eqref{eq:CondICSIds}. Comparing this
expression against Eq.~\eqref{eq:QHall},  the deviation of 
the differential Noether charge constructed by Tachikawa's extension from ours is : 
\begin{equation}
\begin{split}
(\fQNoether)^{IWT}_{_H} 
&= (\fQNoether)_{_H} 
-\bigbr{ \delta \fA \cdot \frac{\partial^2\ICS}{\partial \fF \partial \fF} \cdot  \diffF \fA
+ \delta \fGamma^c{}_b \frac{\partial^2\ICS}{\partial \fR^c{}_b \partial \fR^g{}_h}
  \diffF \fGamma^g{}_h }\\ 
&\,\quad -\bigbr{ \delta \fA \cdot \frac{\partial^2\ICS}{\partial \fF \partial \fR^g{}_h}\  \diffF \fGamma^g{}_h
- \diffF \fA \cdot \frac{\partial^2\ICS}{\partial \fF \partial \fR^g{}_h}\  \delta \fGamma^g{}_h}\\
&\,\quad+d\bigbr{\fZ
+\delta \fA \cdot \frac{\partial^2\ICS}{\partial \fF \partial \fF} \cdot (\Lambda+\ic_\xi\fA)
+ \delta \fGamma^c{}_b \frac{\partial^2\ICS}{\partial \fR^c{}_b \partial \fR^g{}_h}
  \nabla_h \xi^g }\\ 
&\,\quad +d\bigbr{
\delta \fA \cdot \frac{\partial^2\ICS}{\partial \fF \partial \fR^g{}_h}\  \nabla_h \xi^g
+ \delta \fGamma^g{}_h  \frac{\partial^2\ICS}{\partial \fR^g{}_h \partial \fF}
\cdot (\Lambda+\ic_\xi\fA)  } \, . \\
\end{split}
\end{equation}
We note that unlike $(\fQNoether)_{_H}$,  $(\fQNoether)^{IWT}_{_H}$ 
is not covariant. Further 
this non-covariance shows up even if one discards the boundary contributions (which is 
justified when we are interested only in the integral of $\fQNoether$ over a closed surface).
In fact, this non-covariance can be directly traced to the non-covariant terms in the 
pre-symplectic current $(\fPSympl)^{IWT}_{_H}$ in Eq.~\eqref{eq:IWTPSHAllF}. Thus, choosing
a covariant pre-symplectic current $(\fPSympl)_{_H}$ automatically guarantees a 
 $\fQNoether$ which is covariant up to boundary contributions.

%%%%%%%%%%%%%%%%%%%%%%%%
\section{Conclusions and discussions}
\label{sec:Conclusion}
In this paper, we have proposed a new formulation of a differential Noether charge 
for theories in the presence of Chern-Simons terms. Our formulation realizes 
a manifestly covariant pre-symplectic current and differential Noether charge. 
We have also presented a  manifestly covariant derivation of the Tachikawa 
formula for Chern-Simons contribution to entropy. 
When contrasted against Tachikawa's extension that we reviewed 
in \S\ref{sec:Comparison}, our derivation has the additional merit 
of being relatively simple and straightforward.

The critical reader might wonder about the ambiguities in our construction. We have chosen a 
specific pre-symplectic current and a differential Noether charge solely guided by covariance 
and in case of Chern-Simons terms, this is indeed  a stringent constraint which almost uniquely 
determines our choice.  This is in contrast with Tachikawa's extension of the Lee-Iyer-Wald procedure where the ambiguities 
in the definition of the charge are resolved by an explicit prescription which unfortunately 
gives a non-covariant answer for Chern-Simons terms (see \S\ref{sec:Comparison}). 

A more systematic prescription is provided by the Barnich-Brandt-Comp\`{e}re formalism \cite{Barnich:2001jy,
Barnich:2007bf,Compere:2007az} where a particular differential operator (called the homotopy operator)
is constructed to resolve such ambiguities. It would be an interesting test to see whether 
Barnich-Brandt-Compere method gives a covariant  pre-symplectic current and
differential Noether charge for Chern-Simons terms. Given that the homotopy operator is 
itself not manifestly covariant, this would be a highly non-trivial check for Barnich-Brandt-Comp\`{e}re formalism.
Further, a rederivation of our expressions  using the homotopy operator  would then remove
much of the ambiguities in our construction.  An encouraging sign in this 
direction is the fact that for Abelian gauge Chern-Simons terms, our prescription
already agrees with the answer previously derived via the homotopy operator \cite{Compere:2009dp}.

A further advantage to rederiving our construction in the Barnich-Brandt-Comp\`{e}re formalism would be 
the following : it would  then be straightforward to demonstrate that our expression
naturally incorporates algebra of currents in the dual CFT. The current algebra of a 
CFT with anomalies exhibits central terms known as Schwinger terms whose structure 
is completely fixed by the anomaly coefficients. It would be interesting to show that our 
differential Noether charge correctly reproduces this  current algebra structure. Further, 
we might be able to extend to Chern-Simons terms other standard results in the homotopy operator 
formalism. For example, it would be interesting to derive  the generalized Smarr 
relation\cite{Barnich:2004uw} relevant for Chern-Simons terms. A related question 
is whether there is a Wald-like formula for asymptotic charges\cite{Amsel:2012se,Senturk:2012yi} of Chern-Simons terms. 

Another direction in which our results can be generalized is to extend them to $p$-forms
with Green-Schwarz couplings. Green-Schwarz couplings can often be traded for 
Chern-Simons couplings by passing to a  description in terms of a dual $p$-form\cite{Sahoo:2006pm,Tachikawa:2006sz,Sen:2007qy}.
It would be interesting to see whether our method can be used to obtain the same answer without dualizing.

We now turn to a largely unexplored set of questions of much current interest - questions about the 
interplay between anomalies and entanglement entropy. Recently, motivated by the generalized gravitational entropy method\cite{Ryu:2006bv,Lewkowycz:2013nqa}, much progress has been made in understanding how higher-derivative
terms in gravity Lagrangians enter holographic entanglement entropy\cite{Camps:2013zua,Dong:2013qoa}.
However, much of this effort has been focused on covariant Lagrangians and much less is understood about 
Chern-Simons terms (see however \cite{Castro:2014tta} for the case of
the gravitational Chern-Simons term in AdS$_3$). Some of the questions one would like 
answered  in this context are :
\begin{itemize}
\item Can one obtain the entanglement entropy formula for Chern-Simons terms by a dimensional reduction?
If yes, what are the extrinsic curvature correction to the Tachikawa formula? In  AdS$_3$, the authors of 
\cite{Castro:2014tta} have argued that the Tachikawa formula receives no corrections. It would be interesting 
to see whether the same holds in higher dimensions by evaluating Chern-Simons terms on the squashed cone metric.
\item Can one reproduce the bulk Chern-Simons equations of motion  from entanglement entropy \`{a} la \cite{Lashkari:2013koa,Faulkner:2013ica}?
\item If one computes the anomaly contributions to the entanglement entropy equation \cite{Bhattacharya:2012mi,Bhattacharya:2013bna}, are they 
independent of the coupling? These terms would then be the analogue of anomaly-induced terms in hydrodynamics.
\item The structure of anomaly-induced terms in hydrodynamics is captured by a replacement rule\cite{Loganayagam:2012pz,Loganayagam:2012zg} which was recently proved by formal Euclidean methods in \cite{Jensen:2012kj,Jensen:2013rga}. Is there a simpler and a more 
physically transparent proof using anomaly-induced entanglement entropy?
\end{itemize} 

As a first step towards answering these questions, one would first like to check that the expressions proposed in this paper,  when evaluated over 
the fluid/gravity solutions of \cite{Azeyanagi:2013xea} correctly reproduce the anomaly-induced hydrodynamics. That, dear reader, will be the subject of our accompanying paper\cite{Azeyanagi:2014}!

\section*{Acknowledgements}
We would like to thank S.~Bhattacharyya, G.~Comp\`{e}re, S.~Detournay, N.~Iqbal, S.~Minwalla, M.~Rangamani and  Y.~Tachikawa  for valuable discussions.  We would especially like to thank Y.~Tachikawa for various insightful comments and clarifications on the subject of this paper. 
T.~A. is grateful to Universit\'{e} Libre de Bruxelles, Yukawa Institute for Theoretical Physics, 
Institut d'\'{E}tudes Scientifiques de Carg\`{e}se and in particular to Harvard University for hospitality.  
T.~A would like to thank the participants of the YITP workshop ``Holographic vistas on Gravity and Strings". 
T.~A. and G.~N. are grateful to the participants and organizers of the Solvay Workshop on ``Holography for Black Holes and Cosmology".  
We would like to thank the participants and organizers of Strings 2014 in Princeton University and Institute 
for Advanced Study, Princeton.
T.~A. was supported by the LabEx ENS-ICFP: ANR-10-LABX-0010/ANR-10-IDEX-0001-02 PSL*. 
R.~L. was supported by Institute  for Advanced Study, Princeton. 
M.~J.~R. was supported by the European Commission - Marie Curie grant
PIOF-GA 2010-275082.
G.~N. was supported by DOE grant DE-FG02-91ER40654 and the Fundamental Laws Initiative at Harvard.

%%%%%%%%%%%%%%%%HERE APPENDIX%%%%%%%%%%%%%%%%%%%%
\appendix

\section{Detailed computation of \fixform{$(\fQNoether)_{H}$}}
\label{app:QHallDetail}
This Appendix summarizes the detailed derivation of our result for the differential Noether charge in Eq.~\eqref{eq:QbarHall}.
We begin by writing down the Hall part of the pre-symplectic current with the second variation set
equal to the diffeomorphism/gauge variation $\diffF$ generated by $\chi=\{\xi^a,\Lambda\}$~:
\begin{equation}\label{eq:OmegaChiH}
\begin{split}
&(\PSymChi)_{_H}^a \\
&\ =
\half \frac{1}{\sqrt{-G}} \,\delta \brk{
\sqrt{-G}\ \prn{\SpH}^{(bc)a} } \diffF G_{bc} 
- \half \frac{1}{\sqrt{-G}}\, \diffF \brk{
\sqrt{-G}\ \prn{\SpH}^{(bc)a} } \delta G_{bc} \\
&\qquad  + \delta A_e \cdot \prn{\sHFF}^{efa}\cdot  \diffF A_f
+ \delta \Gamma^c{}_{be} \cdot \prn{\sHRR}^{bhefa}_{cg}\cdot  \diffF \Gamma^g{}_{hf} \\ 
&\qquad + \delta A_e \cdot \prn{\sHFR}_g^{hefa}\  \diffF \Gamma^g{}_{hf}
- \diffF A_e \cdot \prn{\sHFR}_g^{hefa}\  \delta \Gamma^g{}_{hf}\, . 
\end{split}
\end{equation}

We will begin by simplifying the first line  in Eq.~\eqref{eq:OmegaChiH} :  
\begin{equation}\label{eq:1stlineOmegaChiH}
\begin{split}
\half &\frac{1}{\sqrt{-G}} \,\delta \brk{
\sqrt{-G}\ \prn{\SpH}^{(bc)a} } \,\diffF G_{bc} 
- \half \frac{1}{\sqrt{-G}} \,\diffF \brk{
\sqrt{-G}\ \prn{\SpH}^{(bc)a} } \delta G_{bc} \\
&\quad= \half \frac{1}{\sqrt{-G}} \,\delta \brk{
\sqrt{-G}\ \prn{\SpH}^{(bc)a} \diffF G_{bc} }  
- \half \frac{1}{\sqrt{-G}} \,\diffF \brk{
\sqrt{-G}\ \prn{\SpH}^{(bc)a} \delta G_{bc}} \, .
\end{split}
\end{equation}
The second term on the right hand side of Eq.~\eqref{eq:1stlineOmegaChiH} evaluates to 
\begin{equation}
\begin{split}
\half &\frac{1}{\sqrt{-G}} \,\diffF \brk{
\sqrt{-G}\ \prn{\SpH}^{(cd)a} \delta G_{cd}}\\ 
&= \half (\nabla_b \xi^b)  \prn{\SpH}^{(cd)a} \delta G_{cd} 
+ \half \xi^b\nabla_b \brk{\prn{\SpH}^{(cd)a} \delta G_{cd}}
- \half (\nabla_b \xi^a)\prn{\SpH}^{(cd)b} \delta G_{cd} \\
&=  -\nabla_b\bigbr{ \half
\brk{ \xi^a \prn{\SpH}^{(cd)b}- \prn{\SpH}^{(cd)a}\xi^b }
\delta G_{cd} }
+ \xi^a \nabla_b\brk{\half  \prn{\SpH}^{(cd)b} \delta G_{cd} }\\
&=  -\nabla_b\bigbr{ \half
\brk{ \xi^a \prn{\SpH}^{(cd)b}- \prn{\SpH}^{(cd)a}\xi^b }
\delta G_{cd} }+ \half \xi^a \delta G_{cd} (\THall)^{cd}\\
&\qquad - \half \xi^a \delta \Gamma^d{}_{cb} \prn{\SpH}^{bc}{}_d \, , 
\end{split}
\end{equation}
where we have used Eq.~\eqref{eq:THdgIntByParts}.
Thus, the first line  in Eq.~\eqref{eq:OmegaChiH} can be written as
\begin{equation} \label{eq:1stlinefinal}
\begin{split}
\half &\frac{1}{\sqrt{-G}} \,\delta \brk{
\sqrt{-G}\ \prn{\SpH}^{(bc)a} } \diffF G_{bc} 
- \half \frac{1}{\sqrt{-G}} \,\diffF \brk{
\sqrt{-G}\ \prn{\SpH}^{(bc)a} } \delta G_{bc} \\
&= \nabla_b\bigbr{ \half
\brk{ \xi^a \prn{\SpH}^{(cd)b}- \prn{\SpH}^{(cd)a}\xi^b }
\delta G_{cd} }- \half \xi^a \delta G_{cd} (\THall)^{cd}\\
&\qquad +\half \xi^a \delta \Gamma^d{}_{cb} \prn{\SpH}^{bc}{}_d 
+\half \frac{1}{\sqrt{-G}}\, \delta \brk{
\sqrt{-G}\ \prn{\SpH}^{(bc)a} \diffF G_{bc} }  \, . 
\end{split}
\end{equation}

After rewriting $(\PSymChi)_{_H}^a$ by using Eq.~\eqref{eq:1stlinefinal}, 
we add to it the term 
$\xi^a (\eom{})_{_H}= (1/2) \xi^a \delta G_{cd} (\THall)^{cd} + \xi^a \delta A_b \cdot \JH^b $
to get 
\begin{equation}\label{eq:OmegaEOMH}
\begin{split}
&(\PSymChi)_{_H}^a + \xi^a (\eom{})_{_H}\\
&= \nabla_b\bigbr{ \half
\brk{ \xi^a \prn{\SpH}^{(cd)b}- \prn{\SpH}^{(cd)a}\xi^b }
\delta G_{cd} }\\
&\qquad + \xi^a \brk{\half\delta \Gamma^d{}_{cb} \prn{\SpH}^{bc}{}_d +\delta A_b \cdot \JH^b }
+\half \frac{1}{\sqrt{-G}} \,\delta \brk{
\sqrt{-G}\ \prn{\SpH}^{(bc)a} \diffF G_{bc} }  \\
&\qquad  + \delta A_e \cdot \prn{\sHFF}^{efa}\cdot  \diffF A_f
+ \delta \Gamma^c{}_{be} \cdot \prn{\sHRR}^{bhefa}_{cg}\cdot  \diffF \Gamma^g{}_{hf} \\ 
&\qquad + \delta A_e \cdot \prn{\sHFR}_g^{hefa}\  \diffF \Gamma^g{}_{hf}
- \diffF A_e \cdot \prn{\sHFR}_g^{hefa}\  \delta \Gamma^g{}_{hf} \, .
\end{split}
\end{equation}

We should subtract from this expression the variation of 
the Hall contribution $\N_{_H}^a$ to the on-shell vanishing Noether current,  which is given by 
\begin{equation}\label{eq:NoetherHallapp}
\begin{split}
\N_{_H}^a &=  \xi_b (\THall)^{ab} + \prn{\Lambda+\xi^c A_c} \cdot \JH^a 
\\
&= \nabla_c \brk{ \half \xi_b\prn{\SpH^{abc}+\SpH^{bac}+\SpH^{cab} } } 
+\half \prn{\nabla_b \xi_c +\nabla_c \xi_b}  \SpH^{(bc)a} \\
&\qquad +\half \nabla_c \xi^b \prn{\SpH}^{ac}{}_b + \prn{\Lambda+\xi^c A_c} \cdot \JH^a \\
&= \nabla_b \brk{ \half \xi_c\prn{\SpH^{acb}+\SpH^{bac}+\SpH^{cab} } } 
+\half  \SpH^{(bc)a} \diffF G_{bc}  \\
&\qquad +\half \nabla_c \xi^d \prn{\SpH}^{ac}{}_d + \prn{\Lambda+\xi^c A_c} \cdot \JH^a \, . \\
\end{split}
\end{equation}
Subtracting the variation of this expression from Eq.~\eqref{eq:OmegaEOMH}, we get
\begin{equation} \label{eq:sumkomareomnoethercurrent}
\begin{split}
&(\PSymChi)_{_H}^a + \xi^a (\eom{})_{_H}-\frac{1}{\sqrt{-G}}\,\delta\brk{\sqrt{-G}\ \N_{_H}^a}\\
&= \nabla_b\Bigl\{\ \half
\brk{ \xi^a \prn{\SpH}^{(cd)b}- \prn{\SpH}^{(cd)a}\xi^b }
\delta G_{cd} \Bigr.\\
&\qquad \qquad \Bigl. -\half \frac{\xi^d}{\sqrt{-G}}\,\delta\brk{\sqrt{-G}\  
 G_{cd}\ \prn{\SpH^{acb}+\SpH^{bac}+\SpH^{cab} } }\ \Bigr\}\\
&\qquad + \xi^a \brk{\half\delta \Gamma^d{}_{cb} \prn{\SpH}^{bc}{}_d +\delta A_b \cdot \JH^b }  \\
&\qquad -\frac{1}{\sqrt{-G}}\,\delta\brk{\sqrt{-G}\ 
\prn{\half \nabla_c \xi^d \prn{\SpH}^{ac}{}_d + \prn{\Lambda+\xi^c A_c} \cdot \JH^a}\ } \\
&\qquad  + \delta A_e \cdot \prn{\sHFF}^{efa}\cdot  \diffF A_f
+ \delta \Gamma^c{}_{be} \cdot \prn{\sHRR}^{bhefa}_{cg}\cdot  \diffF \Gamma^g{}_{hf} \\ 
&\qquad + \delta A_e \cdot \prn{\sHFR}_g^{hefa}\  \diffF \Gamma^g{}_{hf}
- \diffF A_e \cdot \prn{\sHFR}_g^{hefa}\  \delta \Gamma^g{}_{hf} \, .
\end{split}
\end{equation}
Now we want to express the right hand side of the above expression as a total divergence. 
Let us begin by simplifying the first two lines 
outside the divergence in Eq.~\eqref{eq:sumkomareomnoethercurrent} : 
\begin{equation}
\begin{split}
\xi^a &\brk{\half\delta \Gamma^d{}_{cb} \prn{\SpH}^{bc}{}_d +\delta A_b \cdot \JH^b }  \\
&\quad -\frac{1}{\sqrt{-G}}\,\delta\brk{\sqrt{-G}\ 
\prn{\half \nabla_c \xi^d \prn{\SpH}^{ac}{}_d + \prn{\Lambda+\xi^c A_c} \cdot \JH^a}\ } \\
&= \xi^f \brk{\ \half\delta \Gamma^d{}_{cb}
\prn{\delta^a_f  \prn{\SpH}^{bc}{}_d- \delta^b_f \prn{\SpH}^{ac}{}_d}
+ \delta A_b \cdot \prn{\delta^a_f \JH^b - \delta^b_f \JH^a} }\\
&\quad - 
\prn{\half \nabla_c \xi^d 
\frac{1}{\sqrt{-G}}\,\delta\brk{\sqrt{-G}\ \prn{\SpH}^{ac}{}_d} 
+ \prn{\Lambda+\xi^c A_c} \cdot \frac{1}{\sqrt{-G}}\,\delta\brk{\sqrt{-G}\ \JH^a}\ } \\
&= -\prn{\delta \Gamma^d{}_{cb} \prn{\sHRR}_{dg}^{cheab} 
+ \delta A_b \cdot \prn{\sHFR}_g^{heab}} \xi^f R^g{}_{hfe} \\
&\quad -  
\prn{\delta \Gamma^d{}_{cb} \prn{\sHRF}_d^{ceab}+\delta A_b \cdot \prn{\sHFF}^{eab}}
\cdot \xi^f F_{fe}\\
&\quad + \prn{  \nabla_h \xi^g \prn{\sHRR}_{gd}^{hceab} 
+  \prn{\Lambda+\xi^f A_f} \cdot \prn{\sHFR}_d^{ceab} } \nabla_e  \delta \Gamma^d{}_{cb}\\
&\quad + \prn{  \nabla_h \xi^g \prn{\sHRF}_g^{heab} 
+  \prn{\Lambda+\xi^f A_f} \cdot \prn{\sHFF}^{eab} } \cdot \nabla_e  \delta A_b\, ,\\
\end{split}
\end{equation}
where we have used Eqs.~\eqref{eq:HallCond} and \eqref{eq:ixiProp}. Next, we  
shift the covariant derivatives from $\nabla_e  \delta \Gamma^d{}_{cb}$ and 
$\nabla_e  \delta A_b$ by an integration by parts to obtain 
\begin{equation}
\begin{split}
\xi^a &\brk{\half\delta \Gamma^d{}_{cb} \prn{\SpH}^{bc}{}_d +\delta A_b \cdot \JH^b }  \\
&\qquad -\frac{1}{\sqrt{-G}}\,\delta\brk{\sqrt{-G}\ 
\prn{\half \nabla_c \xi^d \prn{\SpH}^{ac}{}_d + \prn{\Lambda+\xi^c A_c} \cdot \JH^a}\ } \\
&\qquad  + \delta A_e \cdot \prn{\sHFF}^{efa}\cdot  \diffF A_f
+ \delta \Gamma^c{}_{be} \cdot \prn{\sHRR}^{bhefa}_{cg}\cdot  \diffF \Gamma^g{}_{hf} \\ 
&\qquad + \delta A_e \cdot \prn{\sHFR}_g^{hefa}\  \diffF \Gamma^g{}_{hf}
- \diffF A_e \cdot \prn{\sHFR}_g^{hefa}\  \delta \Gamma^g{}_{hf}\\
&\quad =
\nabla_e \Bigl\{\ \prn{  \nabla_h \xi^g \prn{\sHRR}_{gd}^{hceab} 
+  \prn{\Lambda+\xi^f A_f} \cdot \prn{\sHFR}_d^{ceab} }  \delta \Gamma^d{}_{cb} \Bigr.\\
&\qquad\,\, \Bigl.+ \prn{  \nabla_h \xi^g \prn{\sHRF}_g^{heab} 
+  \prn{\Lambda+\xi^f A_f} \cdot \prn{\sHFF}^{eab} } \cdot \delta A_b\ \Bigr\}\\
&\quad =
-\nabla_b \Bigl\{\ \prn{  \nabla_h \xi^g \prn{\sHRR}_{gd}^{hcfab} 
+  \prn{\Lambda+\xi^e A_e} \cdot \prn{\sHFR}_d^{cfab} }  \delta \Gamma^d{}_{cf} \Bigr.\\
&\qquad\,\, \Bigl.+ \prn{  \nabla_h \xi^g \prn{\sHRF}_g^{hfab} 
+  \prn{\Lambda+\xi^e A_e} \cdot \prn{\sHFF}^{fab} } \cdot \delta A_f\ \Bigr\}\, . \\
\end{split}
\end{equation}

Combining all the terms together, we finally obtain
\begin{equation}
\begin{split}
-\nabla_b (\QNoether^{\ ab})_{H} 
=(\PSymChi)_{_H}^a + \xi^a (\eom{})_{_H}-\frac{1}{\sqrt{-G}}\,\delta\brk{\sqrt{-G}\ \N_{_H}^a} \, , 
\end{split}
\end{equation}
with 
\begin{equation}\label{eq:QbarHallapp}
\begin{split}
(\QNoether^{\ ab})_{H} 
&= \brk{\nabla_h \xi^g \prn{\sHRR}_{gd}^{hcabf} +
\prn{\Lambda+\xi^e A_e} \cdot \prn{\sHFR}_d^{cabf}\ }\delta \Gamma^d{}_{cf}\\
&\quad +  \brk{ \nabla_h \xi^g \prn{\sHRF}_g^{habf}
+\prn{\Lambda+\xi^e A_e} \cdot \prn{\sHFF}^{abf}\ }\cdot \delta A_f \\
&\quad +\half
\brk{ \prn{\SpH}^{(cd)a}\xi^b- \prn{\SpH}^{(cd)b}\xi^a }
\delta G_{cd}  \\
&\quad +\half \frac{\xi^d}{\sqrt{-G}}\delta\brk{\sqrt{-G}\  
 G_{cd}\ \prn{\SpH^{acb}+\SpH^{bac}+\SpH^{cab} } } \, . 
\end{split}
\end{equation}
%%%%%%%%%%%%%%%%%%%%%%%%%%
\section{Differential forms and Noether charge}
\label{sec:Notationsv1}
In this Appendix, we summarize our notation for the differential forms and present the formulation of the Noether charge in differential forms.

\subsection{Notation : differential forms}\label{ss:forms}

It is often useful to shift to the language of differential forms (denoted by bold
letters in this paper) which is a more efficient way of dealing with antisymmetric tensor indices. In
this Appendix, we summarize our conventions for differential forms.

\begin{itemize}

\item We will denote the volume form on the spacetime by
\begin{equation}\label{eq:volform}
 d^{d+1} x\sqrt{G\ \text{Sign}[G]} = \frac{\text{Sign}[G]}{(d+1)!} \varepsilon_{a_0 a_1\ldots a_{d}} dx^{a_0}\wedge dx^{a_1}\wedge\ldots \wedge dx^{a_{d}} \, ,
\end{equation}
where $G$ denotes the determinant of the metric and $\text{Sign}[G]$ is its signature.

For pseudo-Riemannian metrics describing spacetime, we have $\text{Sign}[G]=-1$ and we take $\varepsilon_{r t  x^1 \ldots x^{d-1} }\equiv  - \sqrt{-G} $ where $r$ is the (spatial) holographic direction with $r\rightarrow \infty$ corresponds to the conformal boundary of AdS$_{d+1}$. The  epsilon tensor for the dual CFT$_{d}$ on ${\mathbb R}^{d-1,1}$ (with the flat metric) is taken to be $\varepsilon_{t x^1 \ldots x^{d-1}}= -1$.

\item We define the Hodge-dual of a $p$-form $\form{V}$ via
\begin{equation}
\label{eq:HodgeDualDef}
(\hodge \form{V})_{a_1 a_2\ldots a_{d+1-p}}
\equiv \frac{\text{Sign}[G]}{p!} V_{b_1 b_2\ldots b_p} \varepsilon^{b_1b_2\ldots b_p}{}_{a_1 a_2\ldots a_{d+1-p}} \, ,
\end{equation}
or, in other words,
\begin{equation}
\hodge \form{V}
\equiv \frac{\text{Sign}[G]}{p!(d+1-p)!}\ V_{b_1 b_2\ldots b_p}\ \varepsilon^{b_1b_2\ldots b_p}{}_{a_1 a_2\ldots a_{d+1-p}}\
dx^{a_1} \wedge dx^{a_2} \ldots \wedge dx^{a_{d+1-p}} \, .
\end{equation}

We note that the definition above is equivalent to
\begin{equation}
 \hodge \prn{ dx_{b_1}\wedge dx_{b_2}\ldots dx_{b_p} }
\equiv \frac{\text{Sign}[G]}{(d+1-p)!}\  \varepsilon_{b_1b_2\ldots b_p a_1 a_2\ldots a_{d+1-p}}\
dx^{a_1} \wedge dx^{a_2} \ldots \wedge dx^{a_{d+1-p}} \, ,
\end{equation}
or
\begin{equation} \label{eq:nonstartostar}
\begin{split}
 dx^{b_1}\wedge dx^{b_2}\ldots dx^{b_p}
 &\equiv \frac{ 1 }{(d+1-p)!}\  \varepsilon^{a_1 a_2\ldots a_{d+1-p}b_1b_2\ldots b_p}\
 \hodge\prn{dx_{a_1} \wedge dx_{a_2} \ldots \wedge dx_{a_{d+1-p}} } \\
&\equiv \frac{  (-1)^{p(d+1-p)} }{(d+1-p)!}\  \varepsilon^{b_1 b_2\ldots b_p a_1 a_2\ldots a_{d+1-p}}\
\hodge\prn{dx_{a_1} \wedge dx_{a_2} \ldots \wedge dx_{a_{d+1-p}} } \, .
\end{split}
\end{equation}

For the boundary CFT$_d$, our convention for the Hodge-dual $\hodgeCFT$ is given by similar expression as in Eq.~\eqref{eq:HodgeDualDef} but with $G_{ab}$ replaced by the flat metric on ${\mathbb R}^{d-1,1}$ and the bulk epsilon tensor replaced by the boundary epsilon tensor  as discussed below Eq.~\eqref{eq:volform}.

\item One of the main uses of Eq.~\eqref{eq:nonstartostar} is in translating expressions of the following form
into components
\begin{equation}
\hodge \form{V} = \form{A}_1 \wedge \form{A}_2 \wedge\ldots \wedge \form{A}_k \, .
\end{equation}
Here $\form{V}$ is a $(d+1-p)$-form , $\form{A}_1 $ is a $q_1$-form, $\form{A}_2 $ is a $q_2$-form etc. such that
$\sum_{i=1}^k\,q_i= p$. We have
\begin{equation}
\begin{split}
\hodge \form{V} &= \form{A}_1 \wedge \form{A}_2 \wedge\ldots \wedge \form{A}_k \\
&= \frac{1}{q_1 ! q_2 ! \ldots q_k !}
(A_1)_{a_1 \ldots a_{q_1}} (A_2)_{b_1\ldots b_{q_2}} \ldots (A_k)_{f_1\ldots f_{q_k}} \\
&\qquad  \qquad dx^{a_1}\wedge \ldots dx^{a_{q_1}}\wedge dx^{b_1}\ldots dx^{b_{q_2}}\wedge \ldots  dx^{f_1}\wedge\ldots
dx^{f_{q_k}}  \\
&= \frac{ 1 }{q_1 ! q_2 ! \ldots q_k !(d+1-p)!}
 \varepsilon^{c_1 c_2\ldots c_{d+1-p} a_1\ldots a_{q_1}b_1\ldots
b_{q_2}\ldots f_1\ldots f_{q_k} } \\
&\qquad  \qquad \
(A_1)_{a_1\ldots a_{q_1}} (A_2)_{b_1\ldots b_{q_2}} \ldots (A_k)_{f_1\ldots f_{q_k}}
\hodge\prn{dx_{c_1} \wedge dx_{c_2} \ldots \wedge dx_{c_{d+1-p}} } \, ,\\
\end{split}
\end{equation}
so that the component of $\form{V}$ is written as 
\begin{equation}\label{eq:starToComponents}
\begin{split}
V^{c_1c_2\ldots c_{d+1-p}} &= \frac{ 1 }{q_1 ! q_2 ! \ldots q_k !
}
 \varepsilon^{c_1 c_2\ldots c_{d+1-p} a_1\ldots a_{q_1} b_1\ldots
b_{q_2}\ldots f_1\ldots f_{q_k} } \\
&\qquad  \qquad \
(A_1)_{a_1\ldots a_{q_1}} (A_2)_{b_1\ldots b_{q_2}} \ldots (A_k)_{f_1\ldots f_{q_k}}\ \, . \\
\end{split}
\end{equation}

\item Given two $p$-forms $\form{V}_1$ and $\form{V}_2$, we have
\begin{equation}
\form{V}_1 \wedge \hodge \form{V}_2 = d^{d+1} x \sqrt{-G}\ \frac{1}{p!}\ (V_1)_{c_1 c_2\ldots c_p} (V_2)^{c_1 c_2\ldots c_p} \, .
\end{equation}

\item Given a $p$-form $\form{V}_1$ and a $q$-form $\form{V}_2$ with $q \geq p$, we have
\begin{equation}
 \form{V}_1 \wedge \hodge \form{V}_2 =
\frac{1}{p!(q-p)!}
\prn{V_1}_{b_1 b_2\ldots b_p} \prn{V_2}^{c_1 c_2\ldots c_{q-p}b_1 b_2\ldots b_p }
\hodge\prn{dx_{c_1} \wedge dx_{c_2} \ldots dx_{c_{q-p}} } \, .
\end{equation}

\item Given a $p$-form $\form{V}$, we introduce a form $\overline{\form{V}}$ such that 
$\form{V} = -\hodge\overline{\form{V}}$. In components, we have
\begin{equation} \label{eq:defofbarnotation}
\begin{split}
(\overline{V})_{a_1 a_2\ldots a_{d+1-p}}
&\equiv -\frac{1}{p!} 
\varepsilon_{a_1 a_2\ldots a_{d+1-p}}{}^{b_1 b_2\ldots b_p} V_{b_1b_2\ldots b_p}\ .
\end{split}
\end{equation}

For a $k$-form $\form{U}$, a result we will need is
\begin{equation}
\begin{split}
\form{U}&\wedge \hodge\overline{\form{V}}\\
&= 
\frac{1}{k!}U_{c_1 c_2\ldots c_k} 
(\overline{V})^{a_1 a_2\ldots a_{d+1-p-k} c_1 c_2\ldots c_k}
\frac{1}{(d+1-p-k)!}\hodge(dx_{a_1}\wedge  \ldots \wedge dx_{a_{d+1-p-k}} )\, . 
\end{split}
\end{equation}
Another result we will need is $\ic_\xi \hodge \form{V} = \hodge\prn{\form{V}\wedge \form{\xi}}$ for any vector $\xi^a$ whose dual one-form is given by $\form{\xi}\equiv G_{ab}\,\xi^a dx^b$.

\end{itemize}

\subsection{Noether charge formalism in differential forms}
It is straightforward to convert our equations about the Noether charge 
formulation to differential forms using the formulae
given in Appendix~\ref{ss:forms}.

We begin by defining the equation of motion form via $\feom{} =-(\eom{})\ \hodge 1$ and the 
pre-symplectic form as the Hodge-dual of the pre-symplectic current by the use of the relation
$\fPSympl = - (\PSympl)^a\  \hodge dx_a$.
All the other forms are defined in a similar fashion following Eq.~\eqref{eq:defofbarnotation}.
The basic equation Eq.~\eqref{eq:divPS} about the divergence of the pre-symplectic current becomes
\begin{equation}\label{eq:divPSf}
d(\fPSympl) = \delta_1 (\feom{2}) - \delta_2 (\feom{1}) \, . 
\end{equation}
We note that various factor of $\sqrt{-G}$ are naturally taken into account in the language of forms.

By introducing the form corresponding to the pre-symplectic potential as $\fPSymplPot{}$, 
Eq.~\eqref{eq:SymplPot} is written as
\begin{equation}\label{eq:deltafPSymplPot}
\fPSympl = -\delta_1 (\fPSymplPot{2}) + \delta_2 (\fPSymplPot{1})\, . 
\end{equation}
Noether's theorem then assumes the form
\begin{equation}
\begin{split}
d\ \hodge \fN &=- \feom{\chi}\ ,\quad   \hodge \fN \simeq 0 \, ,
\end{split}
\end{equation}
while the Komar decomposition Eq.~\eqref{eq:komardecomposition} is of the form
\begin{equation}
\begin{split}
-d\fKomar &= \ic_\xi \fLag - \fPSymplPot{\chi} + \hodge \fN\, . 
\end{split}
\end{equation}
Next, the defining equation Eq.~\eqref{eq:definingeqofnoethercharge} for the differential Noether charge becomes
\begin{equation}
\begin{split}
-d\ (\fQNoether) &= \fPSymChi + \ic_\xi\feom{} + \delta(\hodge \fN)\, .
\end{split}
\end{equation}
Finally, in terms of the Komar charge, the Lee-Iyer-Wald differential Noether charge is (by converting Eq.~\eqref{eq:noetherintermsofkomar} to differential forms):
\begin{equation}
\begin{split}
\fQNoether &= \delta\fKomar - \ic_\xi\fPSymplPot{} \,. 
\end{split}
\end{equation}

\subsection{Einstein-Maxwell contribution}
Here we rewrite the derivation of the Einstein-Maxwell Noether charge in differential forms. 
We first begin with the Lagrangian form for the Einstein-Maxwell theory defined via 
\[ \fLag_{_\text{Ein-Max}} \equiv -\Lag_{_\text{Ein-Max}}\ \hodge 1\, .\]
Thus Eq.~\eqref{eq:LEinMax} becomes
\begin{equation}
\begin{split}
 \fLag_{_\text{Ein-Max}} = \fR^b{}_a\wedge\frac{\hodge (dx^a\wedge dx_b)}{16\pi\GN}
 +\frac{\cc}{8\pi\GN}\hodge 1+ \frac{1}{2\gYM^2} \fF\wedge \hodge \fF \, ,
\end{split}
\end{equation}
where we have introduced Maxwell field strength two-form $\fF\equiv (1/2)F_{ab}\,dx^a\wedge dx^b$ and curvature two-form $\fR^a{}_b \equiv (1/2)R^a{}_{bcd}\,dx^c\wedge dx^d$. Later, we will also use gauge field one-form $\fA\equiv A_a dx^a$ and connection one-form $\fGamma^a{}_b\equiv \Gamma^a{}_{bc}dx^c$. We denote products of curvature two-forms as $(\fR^k)^a{}_b\equiv \fR^a{}_{c_1} \wedge \fR^{c_1}{}_{c_2} \wedge \ldots \wedge \fR^{c_{k-2}}{}_{c_{k-1}}\wedge \fR^{c_{k-1}}{}_b$ and hence $\tr[\fR^k]\equiv (\fR^k)^a{}_a$ is understood as a matrix-trace.

We remind the reader that given our orientation convention in AdS, we have 
$\hodge 1 = -\sqrt{-G}\ d^{d+1}x$ and hence the Einstein-Maxwell action is given by 
\[S_{_\text{Ein-Max}}=  \int \fLag_{_\text{Ein-Max}} = \int \brk{
 \fR^b{}_a\wedge\frac{\hodge (dx^a\wedge dx_b)}{16\pi\GN}
 +\frac{\cc}{8\pi\GN}\hodge 1+ \frac{1}{2\gYM^2} \fF\wedge \hodge \fF }\, . 
\]
The corresponding pre-symplectic potential in Eq.~\eqref{eq:PSymplPotEinMax} becomes
\begin{equation}
\begin{split}
(\fPSymplPot{})_{_\text{Ein-Max}} &\equiv -(\PSymplPot{})_{_\text{Ein-Max}}^a\ \hodge dx_a\\
&=    \delta\fGamma^b{}_a\  \frac{\hodge(dx^a\wedge dx_b)}{16\pi \GN} 
+\delta \fA  \cdot  \frac{\hodge \fF}{\gYM^2}   \\
&=    \delta\fGamma^b{}_a\  \frac{\partial \fLag_{_\text{Ein-Max}} }{\partial \fR^b{}_a} 
+\delta \fA  \cdot   \frac{\partial \fLag_{_\text{Ein-Max}} }{\partial \fF}   \, , 
\end{split}
\end{equation}
and the Hodge-dual of the pre-symplectic current in Eq.~\eqref{eq:PSEinMAx} is 
\begin{equation}
\begin{split}
(\fPSympl)_{_\text{Ein-Max}} &\equiv  -(\PSympl)_{_\text{Ein-Max}}^a\ \hodge dx_a\\
&=\brk{ \   \delta_1\fGamma^b{}_a\  \delta_2\prn{\frac{\hodge(dx^a\wedge dx_b)}{16\pi \GN} }
+\delta_1\fA  \cdot  \delta_2\prn{\frac{\hodge \fF}{\gYM^2}}   
}  \\
&\qquad - \brk{ \   \delta_2\fGamma^b{}_a\  \delta_1\prn{\frac{\hodge(dx^a\wedge dx_b)}{16\pi \GN} }
+\delta_2\fA  \cdot  \delta_1\prn{\frac{\hodge \fF}{\gYM^2}}   
} \, . 
\end{split}
\end{equation}

Moving on to the Komar charge, we have
\begin{equation}
\begin{split}
(\fKomar)_{_\text{Ein-Max}} 
&\equiv - \frac{1}{2!} (\Komar)^{ab}_{_\text{Ein-Max}}\ \hodge(dx_a\wedge dx_b)\\
& =
\nabla_a \xi^b\  \frac{\hodge(dx^a\wedge dx_b)}{16\pi \GN} 
+\prn{\Lambda+\ic_\xi \fA}  \cdot  \frac{\hodge \fF}{\gYM^2} \\  
&=    \nabla_a \xi^b\  \frac{\partial \fLag_{_\text{Ein-Max}} }{\partial \fR^b{}_a} 
+\prn{\Lambda+\ic_\xi \fA}  \cdot   \frac{\partial \fLag_{_\text{Ein-Max}} }{\partial \fF}   \, . 
\end{split}
\end{equation}

The Einstein-Maxwell contribution to the differential Noether charge in Eq.~\eqref{eq:QbarEinMax}
can be written in terms of forms as
\begin{equation}
\begin{split}
(&\fQNoether)_{_\text{Ein-Max}} \\
&\ \equiv -\frac{1}{2!} (\QNoether)^{ab}_{_\text{Ein-Max}}\ \hodge(dx_a\wedge dx_b)\\
&\ =
\delta\brk{ \   \nabla_a \xi^b\  \frac{\hodge(dx^a\wedge dx_b)}{16\pi \GN} 
+\prn{\Lambda+\ic_\xi \fA}  \cdot  \frac{\hodge \fF}{\gYM^2}   
}\\
&\qquad -
\ic_\xi \brk{ \   \delta\fGamma^b{}_a\  \frac{\hodge(dx^a\wedge dx_b)}{16\pi \GN} 
+\delta \fA  \cdot  \frac{\hodge \fF}{\gYM^2}   
}  \, , 
\end{split}
\end{equation}
which can be simplified further to give 
\begin{equation}\label{eq:QEinMax}
\begin{split}
(&\fQNoether)_{_\text{Ein-Max}} \\
&\ =
 \ \half\  \nabla_a \xi^b\ \delta\brk{ \frac{\hodge(dx^a\wedge dx_b)}{8\pi \GN} }
+\prn{\Lambda+\ic_\xi \fA}  \cdot  \delta\brk{ \frac{\hodge \fF}{\gYM^2}   }
\\
&\qquad +
 \half\  \delta\fGamma^b{}_a\  \frac{\ic_\xi\hodge(dx^a\wedge dx_b)}{8\pi \GN} 
+\delta \fA  \cdot  \frac{\ic_\xi \hodge \fF}{\gYM^2} \, .   
\end{split}
\end{equation}

\subsection{Hall contribution}
As the next step, we rewrite the derivation of the Hall contribution to the Noether charge in differential forms. 
We start with the defining equation for
the Hall conductivities in Eq.~\eqref{eq:HallCond} which can be 
stated in terms of forms as
\begin{equation}
\begin{split}
 -\delta\prn{\hodge \fJH } 
&\equiv   \delta \fF \cdot \fsHFF + \delta \fR^g{}_{h} \wedge \prn{\fsHFR}_g^{h} \, ,  \\
-\half  \delta  \prn{\hodge \fSpH}^{b}{}_c 
&\equiv \delta \fF \cdot \prn{\fsHRF}^{b}_c   + 
  \delta \fR^g{}_{h} \wedge \prn{\fsHRR}_{cg}^{bh} \, .\\
\end{split}
\end{equation}
We can now use the expression for the Hall currents
\begin{equation}
\begin{split}
-\hodge \fJH &\equiv \frac{\partial \fPCFT}{\partial \fF}\ ,\quad
-\half  \prn{\hodge \fSpH}^{b}{}_c 
\equiv \frac{\partial \fPCFT}{\partial \fR^c{}_b} \, , 
\end{split}
\end{equation}
to get the generalized Hall conducetivities
\begin{equation}\label{eq:fHallCond}
\begin{split}
 \fsHFF \equiv \frac{\partial^2 \fPCFT}{\partial \fF \partial \fF}\ &,\
\prn{\fsHRR}_{ch}^{bg} \equiv \frac{\partial^2 \fPCFT}{\partial \fR^c{}_b \partial \fR^h{}_g}\ ,\\ 
\prn{\fsHFR}_h^{g} \equiv &\prn{\fsHRF}^{g}_h \equiv  
\frac{\partial^2 \fPCFT}{\partial \fF \partial \fR^h{}_g} \, . 
\end{split}
\end{equation}

To restate the property Eq.~\eqref{eq:ixiProp}, we first rewrite it by contracting with an 
arbitrary vector $\xi^f$ :
\begin{equation}\label{eq:ixiPropContracted}
\begin{split}
  \JH^a \xi^b -  \JH^b \xi^a
&=  \prn{\sHFF}^{eab}\cdot  \xi^f F_{fe}+ 
  \prn{\sHFR}_g^{heab}\  \xi^f R^g{}_{hfe} \, ,\\
  \half (\SpH)^{ac}{}_d \xi^b - \half (\SpH)^{bc}{}_d \xi^a
&=  \prn{\sHRF}^{ceab}_d \cdot \xi^f F_{fe}  + 
 \prn{\sHRR}_{dg}^{cheab}\  \xi^f R^g{}_{hfe} \, .  \\
\end{split}
\end{equation}
We can now multiply both sides by $-(1/2) \hodge(dx_a\wedge dx_b)$ and use 
$\hodge(\form{V}\wedge\form{\xi})=\ic_\xi\hodge\form{V}$ for an arbitrary form $\form{V}$ to get 
\begin{equation}\label{eq:ixiPropF}
\begin{split}
  -\ic_\xi\hodge \fJH
&=  \fsHFF \cdot  \ic_\xi\fF + 
  \prn{\fsHFR}_g^{h}\  \ic_\xi\fR^g{}_{h} \, ,  \\
   -\ic_\xi\half(\hodge\fSpH)^{c}{}_d
&=  \prn{\fsHRF}^{c}_d \cdot \ic_\xi\fF  + 
 \prn{\fsHRR}_{dg}^{ch}\  \ic_\xi\fR^g{}_{h} \, ,\\
\end{split}
\end{equation}
or 
\begin{equation}\label{eq:ixiPropF2}
\begin{split}
  \ic_\xi\frac{\partial \fPCFT}{\partial \fF}
&=  \frac{\partial^2 \fPCFT}{\partial \fF \partial \fF}\cdot  \ic_\xi\fF + 
   \frac{\partial^2 \fPCFT}{\partial \fF \partial \fR^g{}_{h}}\  \ic_\xi\fR^g{}_{h} \, , \\
   \ic_\xi\frac{\partial \fPCFT}{\partial \fR^d{}_c}
&=  \frac{\partial^2 \fPCFT}{\partial \fR^d{}_c \partial \fF } \cdot \ic_\xi\fF  + 
 \frac{\partial^2 \fPCFT}{\partial \fR^d{}_c \partial \fR^g{}_{h}}\  \ic_\xi\fR^g{}_{h} \, , \\
\end{split}
\end{equation}
which is just the statement that the operator $\ic_\xi$ acts as a derivation.

Next, the pre-symplectic current in Eq.~\eqref{eq:PSHAll} becomes
\begin{equation}\label{eq:PSHAllF}
\begin{split}
(\fPSympl)_{_H}
&\ = -(\PSympl)_{_H}^a\ \hodge dx_a  \\
&\ =
-\half \delta_1 \brk{
 \prn{\SpH}^{(bc)a} \ \hodge dx_a } \delta_2 G_{bc} 
+ \half  \delta_2 \brk{
\prn{\SpH}^{(bc)a}\ \hodge dx_a } \delta_1 G_{bc} \\
&\quad  + \delta_1 \fA \cdot \fsHFF \cdot  \delta_2 \fA
+ \delta_1 \fGamma^c{}_b \cdot \prn{\fsHRR}^{bh}_{cg}\cdot  \delta_2 \fGamma^g{}_h \\ 
&\quad + \delta_1 \fA \cdot \prn{\fsHFR}_g^{h}\  \delta_2 \fGamma^g{}_h
- \delta_2 \fA \cdot \prn{\fsHFR}_g^h\  \delta_1 \fGamma^g{}_h\, . 
\end{split}
\end{equation}

Finally, the expression for Hall contribution to the differential Noether charge
given in Eq.~\eqref{eq:QbarHallapp} becomes 
\begin{equation}\label{eq:QHall}
\begin{split}
(\fQNoether)_{H} 
&=  \delta \fGamma^d{}_{c} \wedge\brk{\nabla_h \xi^g \prn{\fsHRR}_{gd}^{hc} +
\prn{\Lambda+ \ic_\xi \fA} \cdot \prn{\fsHFR}_d^{c}\ }\\
&\quad + \delta \fA \cdot \brk{ \nabla_h \xi^g \prn{\fsHRF}_g^{h}
+\prn{\Lambda+\ic_\xi \fA} \cdot \prn{\fsHFF}\ } \\
&\quad -\half  \delta G_{cd}
\prn{\SpH}^{(cd)a}\ic_\xi \hodge dx_a  \\
&\quad - \xi^d\delta\brk{  
 \half G_{cd}\ \prn{\SpH^{acb}+\SpH^{bac}+\SpH^{cab} }
 \frac{1}{2!} \hodge(dx_a \wedge dx_b) } \, . \\ 
\end{split}
\end{equation}

\bibliographystyle{utphys}
\bibliography{covNoetherCS}

\end{document}